      [1999/12/01 v1.4c Il Nuovo Cimento]
\documentclass[varenna]{cimento}

\usepackage{graphicx}  
\usepackage{bm}
\def\la{\mathrel{\mathpalette\fun <}}
\def\ga{\mathrel{\mathpalette\fun >}}
\def\fun#1#2{\lower3.6pt\vbox{\baselineskip0pt\lineskip.9pt
  \ialign{$\mathsurround=0pt#1\hfil##\hfil$\crcr#2\crcr\sim\crcr}}}
\def\sss{{\hbox{\boldmath$\sigma$}}}
\def\nab{{\hbox{\boldmath$\nabla$}}}

\def\bbeta{{\hbox{\boldmath$\beta$}}}

\title{High Energy Neutrinos and Cosmic Rays}

\author{G\"unter Sigl}

\institute{II. Institut f\"ur theoretische Physik, Universit\"at Hamburg,
Luruper Chaussee 149, D-22761 Hamburg, Germany}

\shortauthor{G.~Sigl}

\begin{document}

\maketitle

\begin{abstract}
This is a summary of a series of lectures on the current experimental
and theoretical status of our understanding of origin and nature of
cosmic radiation. Specific focus is put on ultra-high energy cosmic radiation above
$\sim10^{17}\,$eV, including secondary neutral particles and in
particular neutrinos. The most important open questions are related to
the mass composition and sky distributions of these particles as well as on the
location and nature of their sources. High energy neutrinos at GeV
energies and above from extra-terrestrial sources have not yet been
detected and experimental upper limits start to put strong contraints
on the sources and the acceleration mechanism of very high energy
cosmic rays.
\end{abstract}

\maketitle


\section{Galactic and Extragalactic Primary Cosmic Radiation: A Short
  Overview}

\begin{figure}[ht]
\includegraphics[width=0.5\textwidth,clip=true,angle=0]{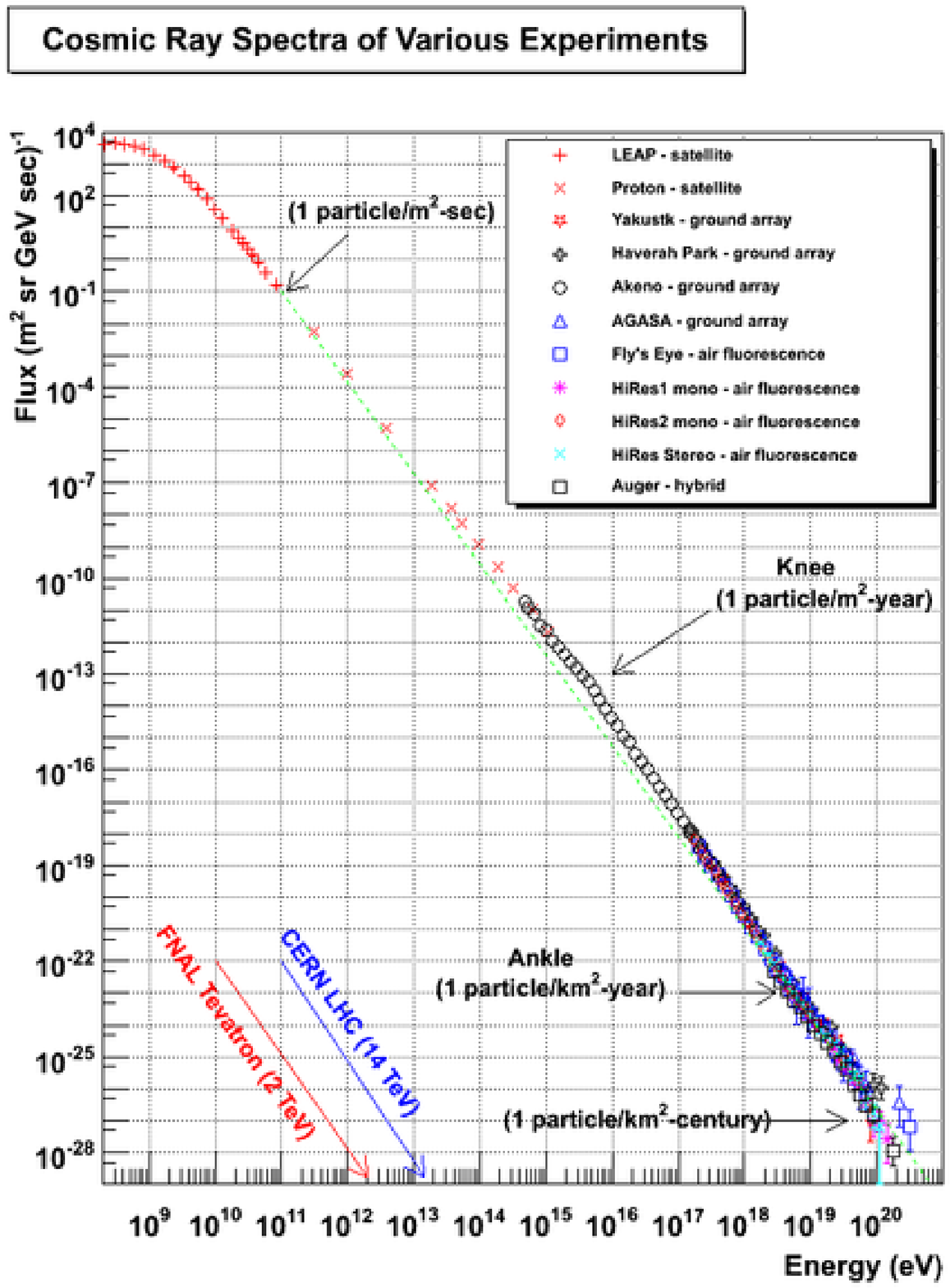}
\includegraphics[width=0.5\textwidth,clip=true,angle=0]{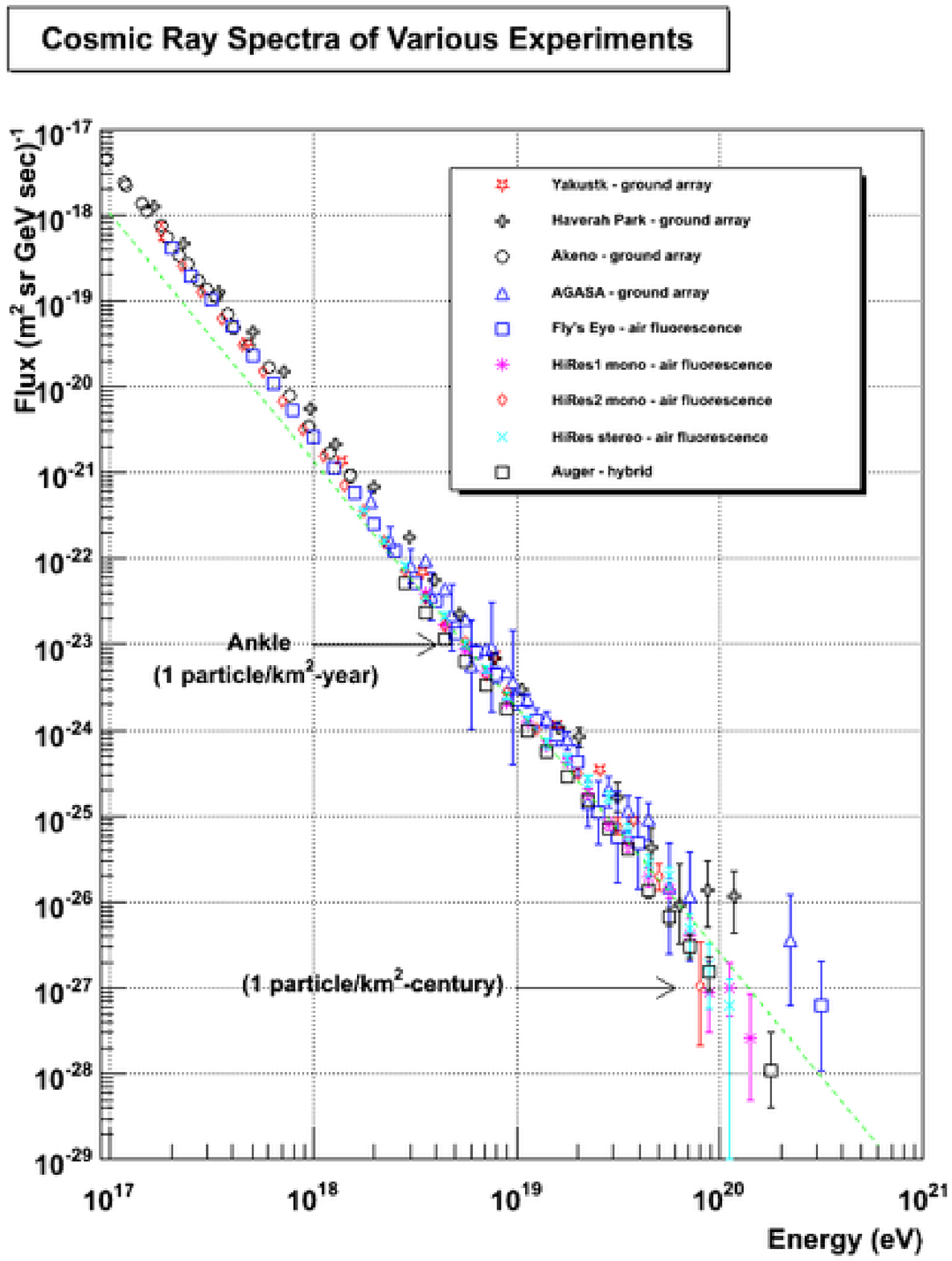}
\caption[...]{The all-particle cosmic ray spectrum and experiments
  relevant for its detection. The right panel is a blow-up of the
  highest energy region. Taken from {\sl http://www.physics.utah.edu/~whanlon/spectrum.html}.}
\label{fig:cr-spectrum}
\end{figure}

High energy cosmic ray (CR) particles reach from energies below 100
MeV up to at least several $10^{20}\,$eV. One of the open questions is
whether the spectrum continues to even higher energies and we just
have not been able to detect it because of limited statistics or has a
hard cut-off for principal physical reasons. Over the observed energy range, the
differential all-particle flux drops by some 32 orders of magnitude,
see Fig.~\ref{fig:cr-spectrum}. Cosmic rays interact in Earth's
atmosphere and are thus shielded revealing their existence on the
ground only by indirect effects such as ionization and the formation
of showers of secondary charged particles. At energies below a few
hundred GeV these showers die out high above the atmosphere and
are not sufficiently luminous to be visible from the ground such that
they can only be detected with balloons or from satellites. Above a
few hundred GeV they can be detected with telescopes from the ground
and above $\sim10^{15}\,$eV, depending on the altitude of the
detector, the charged secondaries and muons are sufficiently numerous
to be detected on the ground. At the highest
energies these air showers cover areas up
to many km$^2$. In 1912 Victor Hess discovered CRs by measuring ionization from
a balloon~\cite{hess}, and in 1938 Pierre Auger proved the existence of
extensive air showers (EAS) caused by primary particles
with energies above $10^{15}\,$eV by simultaneously observing
the arrival of secondary particles in Geiger counters many meters
apart~\cite{auger_disc}.

Still, after almost 90 years of research, the origin of cosmic rays
is largely an open question~\cite{crbook}: Only for particles of
kinetic energy below 100 MeV we are sure that they have to come from
the Sun the solar wind shields charged particles coming
from outside the solar system at such energies. At higher energies the
CR spectrum exhibits little structure and is approximated
by broken power laws $\propto E^{-\gamma}$:
At the energy $E\simeq4\times 10^{15}\,$eV
called the ``knee'', the flux of particles per area, time, solid angle,
and energy steepens from a power law index $\gamma\simeq2.7$
to one of index $\simeq3.0$. The bulk of the CRs up to at least
that energy is believed to originate within the Milky Way Galaxy,
typically by shock acceleration in supernova remnants. One of the main
arguments for this comes from a consideration of energy budgets which
we will consider first.

If the CR accelerators are Galactic, they must replenish the CR flux
that escapes from the Galaxy in order to sustain the observed Galactic CR differential
intensity $j(E)$. Their total luminosity in CR must therefore satisfy
$L_{\rm CR}=(4\pi/c)\int dEdV t_{\rm CR}(E)^{-1}Ej(E)$, 
where $t_{\rm CR}(E)$ is the mean residence time of CR
with energy $E$ in the Galaxy and $V$ is the volume. $t_{\rm
CR}(E)$ can be estimated from the mean column
density, $X(E)$, of gas in the interstellar medium that
Galactic CR with energy $E$ have traversed. Interaction of the primary CR
particles with the gas in the interstellar medium leads to production of
various secondary species. From the secondary to
primary abundance ratios of Galactic CR it was infered
that~\cite{swordy}
\begin{equation}
X(E)=\rho_g t_{\rm CR}(E)\simeq 6.9
\left(\frac{E}{20Z\,{\rm GeV}}\right)^{-0.6}\,{\rm g}\,{\rm cm}^{-2}\,,
\end{equation}
where $\rho_g$ is the mean density of interstellar gas and $Z$ is the
mean charge number of the CR particles.
The mean energy density of CR and the total mass of gas in the Milky Way
that have been inferred from the diffuse Galactic $\gamma-$ray,
X-ray and radio emissions are $u_{\rm CR}=(4\pi/c)\int dE Ej(E)\simeq
1\,{\rm eV}\,{\rm cm}^{-3}$
and  $M_{g}\sim\rho_g V \sim 4.8\times 10^9M_\odot$,
respectively. Hence, simple integration yields 
\begin{equation}
L_{\rm CR}\sim M_{g}\int dE\frac{Ej(E)}{X(E)}
\sim 1.5\times10^{41}\,{\rm erg}\,{\rm sec}^{-1}\,.\label{lumcr}
\end{equation}
This is about 10\% of the estimated total power output in the form
of kinetic energy of the ejected material in    
Galactic supernovae which, from the energetics point of view,
could therefore account for most of the CR. We note
that the energy release from other Galactic sources,
e.g. ordinary stars or isolated neutron
stars~\cite{bbdgp} is expected to be too small, even for
UHECR. Together with other considerations (see section~\ref{section4}
below) this leads to the widely held notion that CR at least up to the knee
predominantly originate from first-order Fermi acceleration
in supernova remnants.

Another interesting observation is that the energy density in the form of 
CR is comparable both to the energy density in the Galactic
magnetic field ($\sim10^{-6}\,$G) as well as that in the
turbulent motion of the gas,
\begin{equation}
  u_{\rm CR}\sim\frac{B^2}{8\pi}\sim\frac{1}{2}\,\rho_g v_t^2
  \,,\label{crb}
\end{equation}
where $\rho_g$ and $v_t$ are the density and turbulent velocity
of the gas, respectively.
This can be expected from pressure equilibrium between the
(relativistic) CR, the magnetic field, and the gas flow. If
Eq.~(\ref{crb}) roughly holds not only in the Galaxy but also
throughout extragalactic space, then we would expect the
extragalactic CR energy density to be considerably
smaller than the Galactic one which is another argument in favor
of a mostly Galactic origin of the CR observed near
Earth. We note, however, that, in order for
Eq.~(\ref{crb}) to hold, typical CR diffusion time-scale over
the size of the system under consideration must be smaller than
its age. This is not the case, for example, in clusters of
galaxies if the bulk of CR are produced in the member galaxies or in 
cluster accretion shocks.

Above the knee, the all-particle CR spectrum continues with a further
steepening to $\gamma\simeq3.3$
at $E\simeq4\times 10^{17}\,$eV, sometimes called the ``second knee''.
There are experimental indications that the mass composition
changes from light, mostly protons, at the knee to domination by
iron and even heavier nuclei at the second knee~\cite{Hoerandel:2004gv}.
This is in fact expected in any scenario which is dominated by
propagation in magnetic fields, both during acceleration and
propagation from the source to the observer, and energy losses can be
neglected, such that particle transport only depends
on rigidity, the ratio of energy to charge, $E/Z$. This is usually
true for baryonic Galactic cosmic rays because their interaction
probability during the lifetime of the Galaxy is less than unity.
Energy losses and interactions will in general break the degeneracy
between $E$ and $Z$ because they will depend on $E$ and $Z$ (and
possibly other quantities, such as atomic number $A$) separately. They
become important for extra-galactic cosmic ray propagation at
ultra-high energies above $\sim10^{19}\,$eV.

Above the so called ``ankle'' or ``dip'' at $E\simeq5\times10^{18}\,$eV, the
spectrum flattens again to a power law of index $\gamma\simeq2.8$.
This is often interpreted as a cross over from a Galactic
component to a harder component of extragalactic
origin. The Galactic component may steepen or cut off completely
because cosmic rays produced within our Galaxy are not confined by 
the Galactic magnetic field any more or because Galactic sources or
more limited in terms of their maximal acceleration energy than
extragalactic sources. However, it is also possible that the
extra-galactic component already starts to dominate below
the ankle, for example, around the second-knee~\cite{Aloisio:2006wv}
at a few times $10^{17}\,$eV. In this case, the dip at
$E\simeq5\times10^{18}\,$eV could also be explained as a feature
induced by pair production of the extragalactic cosmic rays, provided
they are predominantly protons. The observed feature is well fit for a
relatively steep injection spectrum $\propto E^{-2.6-2.7}$. Below a few times $10^{17}\,$eV
diffusion in extra-galactic magnetic fields (EGMF) induces an energy
dependent horizon beyond which extragalactic sources become unobservable at Earth
because diffusion times become larger than the age of the Universe~\cite{Lemoine:2004uw}.
In addition, the effective volume-averaged injection spectrum has to
flatten somewhere below $\sim10^{18}\,$eV to something not much
steeper than $E^{-2}$, otherwise the power going into low energy
cosmic rays would become prohibitive and $pp$ interactions with the ambient gas
would produce a flux of GeV--TeV $\gamma-$rays that would be higher
than observed.

The proton dominance around the dip required in the low cross-over
scenario can be achieved either because preferentially 
protons are accelerated or because extended EGMF lead to strong
photo-spallation during propagation~\cite{Sigl:2005md}.
Experimental information on the mass composition above
$\simeq10^{17}\,$eV is sparse~\cite{Watson:2004ew}. The situation is
particularly inconclusive around $10^{18}\,$eV where the
HiRes~\cite{Abbasi:2004nz} and
HiRes-MIA~\cite{Abu-Zayyad:2000ay} data suggest a light (proton
dominated) composition, whereas other experiments indicate a heavy
composition~\cite{Hoerandel:2004gv}. Above $\sim10^{17}\,$eV the
inferred mass composition increasingly depends on extrapolations
of hadronic interaction into an energy range which cannot be measured
directly because the center of mass energy of a cosmic ray of energy
$E$ with a proton of mass $m_p$, $s^{1/2}\simeq(2m_p
E)^{1/2}\simeq40\,(E/10^{18}\,{\rm eV})^{1/2}\,$TeV, surpasses the maximal energy
that can be attained in the laboratory. This makes them accordingly
uncertain~\cite{Watson:2004ew}. The mass composition could become
lighter again above the ankle, although a significant heavy component is not
excluded either.

Independently of mass composition, it is clear that cosmic rays
above the ankle should have an extragalactic origin because they show
no correlation with the Galactic disc. Would they
have a dominantly galactic origin, they would show an anisotropy toward the
galactic plane since the gyro radius of a cosmic ray of energy $E$ and
charge $eZ$ in a magnetic field of strength $B$, $r_g\simeq
E/(eZB)\simeq1\,(E/10^{18}\,{\rm eV})/Z/(B/\mu{\rm G})$ becomes
larger than the scale height of the Galactic disc and galactic magnetic fields can no longer
isotropize the cosmic rays. Cosmic rays above $10^{18}\,$eV are
usually called ultra-high energy cosmic rays (UHECRs).

\begin{figure}[ht]
\includegraphics[width=0.9\textwidth,clip=true,angle=0]{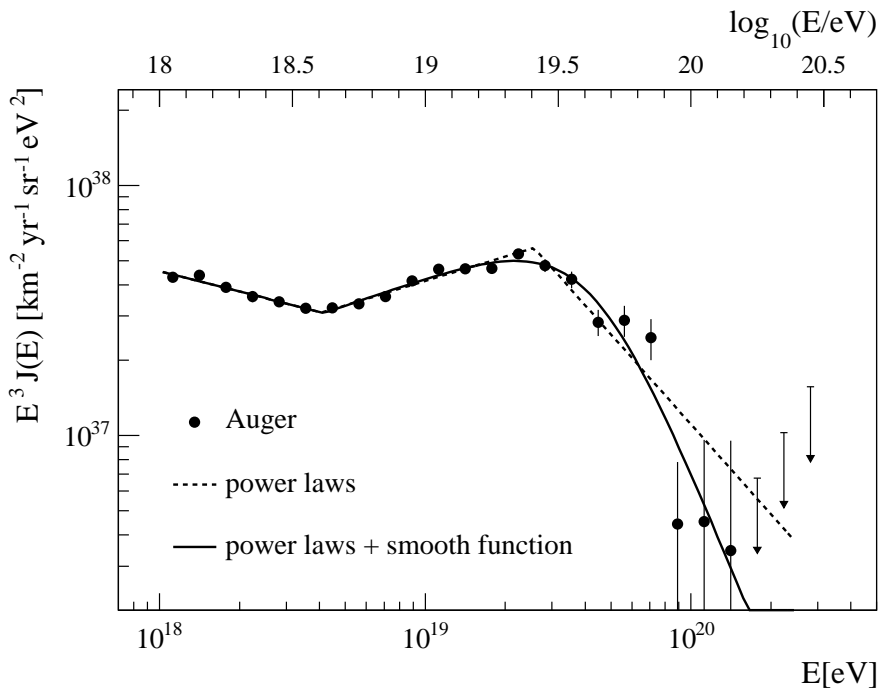}
\caption[...]{The all-particle cosmic ray spectrum measured by the
  Pierre Auger Observatory, from Ref.~\cite{auger:2011pj}.}
\label{fig:auger-spectrum}
\end{figure}

The highest energy cosmic rays in particular
have challenged the imagination of physicists and astrophysicists alike.
The first cosmic ray with energy above $10^{20}\,$eV was discovered by
John Lindsley in 1963 at the Volcano Ranch Observatory~\cite{Linsley:1963km}.
In the 90s the famous ``Fly's Eye event'' with an energy
$\simeq3\times10^{20}\,$eV~\cite{Bird:1994uy} was observed and quickly
scientists were starting to look for astronomical sources~\cite{Elbert:1994zv}.
Around the same time, the Akeno Giant Air Shower Array (AGASA)
observed an UHECR spectrum continuing seemingly as a power law around
$10^{20}\,$eV. This was contrary to expectations and caused excitement because the famous
Greisen-Zatsepin-Kuzmin (GZK) effect~\cite{gzk} predicts that nucleons loose
most of their energy within about 20 Mpc above a threshold of
$\simeq6\times10^{19}\,$eV~\cite{stecker} due to pion production on
the cosmic microwave background (CMB). This energy loss is unavoidable 
because the CMB is a relic of the early Universe and is thus
all-pervading. Iron nuclei have a similar reach above
$\simeq6\times10^{19}\,$eV and intermediate mass nuclei have an even
smaller horizon~\cite{Harari:2006uy,Allard:2008gj}. This implies that above this so-called GZK threshold only sources
within $\simeq50\,$Mpc should be visible. As long as there is no strong
over-density of UHECR sources within that distance scale, this would
predict a strong suppression of the UHECR flux above the GZK
threshold. This is often, somewhat misleadingly, called
the ``GZK cutoff''. It is, however, not a strict cut-off because
sources within $\simeq50\,$Mpc are still visible up to much higher
energies. Meanwhile, a flux suppression consistent with the
GZK effect has been observed by the more recent High Resolution Fly's
Eye~\cite{hires-spec,hires} and Pierre
Auger~\cite{auger-spec,auger:2011pj,pierre_auger} experiments,
see Fig.~\ref{fig:auger-spectrum}. It is likely that the seeming
absence of the GZK suppression in the AGASA spectrum was due energy
calibration problems.

Excellent recent and more in-depth reviews on galactic and
extragalactic cosmic rays can be found in
Ref.~\cite{Castellina:2011gn} and~\cite{LetessierSelvon:2011dy}, respectively.

\section{Cosmic Ray Acceleration}
\label{section2}
\subsection{Shock Acceleration}
In 1949 Enrico Fermi observed that when a charged particles collides
agains moving magnetic fields, for example magnetized interstellar
clouds moving in random directions with a velocity $v$ within our
Galaxy, on average there is an average fractional energy gain per
collision of order $v^2$~\cite{Fermi:1949ee}. This is because although the
particle gains energy when moving towards the ``magnetic mirror'' and
looses when moving away from it, on average the probability for
approaching the mirror is higher than the one for receding from the
mirror. Nowadays this is called ``second order Fermi
acceleration''. However, if the ``magnetic mirror'' is regularly
shaped, for example, as a plain wave as is the case at a magnetized
shock front, the fractional energy gain per reflection turns out to be
first order in the velocity $v$ of the moving mirror, which strongly
increases the acceleration efficiency for non-relativistic
motion. This case today is known as ``first order Fermi acceleration''
and below we will focus on this case which we will discuss in detail.

We start with a few general considerations. The force of an
electromagnetic field ${\bf E},{\bf B}$ on a particle of charge $eZ$
and velocity $\bbeta$ is given by
\begin{equation}\label{eq:F_EM}
  {\bf F}=eZ\left({\bf E}+\bbeta\times{\bf B}\right)\,.
\end{equation}
Consider a particle of momentum ${\bf p}$ and energy $E$ gyrating with an angular
frequency $\omega_g$ in a homogeneous magnetic field ${\bf B}_0$.
It will experience a force $\omega_g{\bf p}=eZ({\bf p}\times{\bf B}_0)/E$,
resulting in
\begin{equation}\label{eq:omega_g}
  \omega_g=\frac{eZB_\perp}{E}\,,
\end{equation}
where  $B_\perp=|{\bf p}\times{\bf B}_0|/p$ is the modulus of the
component of ${\bf B}_0$ perpendicular to the motion of the charge.
The gyro-radius, which is also called Larmor radius, is given by
\begin{equation}\label{eq:gyro}
  r_g(p)=\frac{\beta}{\omega_g}=\frac{p/E}{\omega_g}=\frac{p}{eZB_\perp}\,.
\end{equation}
Note that the equation of motion only depends on rigidity, $p/Z$.

Astrophysical plasmas are often highly conducting, thus when the plasma
flow velocity is ${\bf v}$, we have ${\bf E}\simeq-{\bf v}\times{\bf B}$,
and thus ${\bf E}=0$ in the plasma rest frame.

Charged particles with a gyro-radius Eq.~(\ref{eq:gyro}) much smaller
than the size of the magnetized region are diffusing in the magnetic
field. In a general diffusion process a particle with a mean scattering
length $\lambda$ over which it changes direction will propagate an average
squared distance $\langle d^2\rangle\sim \lambda\,t$ over time $t$. The in
general energy dependent scattering length $\lambda(p)$ is often called the
diffusion coefficient $D(p)$. In case of diffusion in magnetic fields,
charged particles typically scatter on inhomogeneities of the magnetic
field and energy loss due to collisions with ambient gas or low energy
photons is often negligible, except at the highest energies. In this case
the diffusion coefficient $D(p)$ will depend on the detailed structure of
the magnetic field, but will typically be larger than or comparable to
the gyro-radius Eq.~(\ref{eq:gyro}),
\begin{equation}\label{eq:bohm}
  D(p)\ga\frac{1}{3}r_g(p)\,.
\end{equation}
The lower limit is often called the {\it Bohm limit}.

\begin{figure}[ht]
\includegraphics[height=0.6\textwidth,clip=true,angle=0]{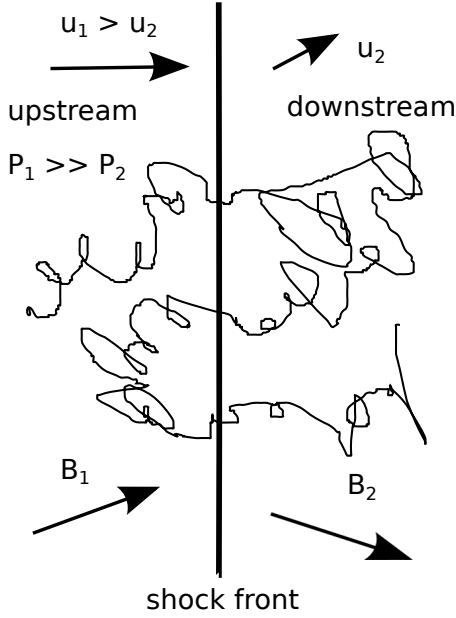}
\caption[...]{For a plane, adiabatic shock the jumps in mass density,
velocity, and pressure are given by the Rankine-Hugoniot conditions
and determine the compression ratio $r\equiv\rho_2/\rho_1$. Upstream and
downstream regions are denoted with index 1 and 2, respectively.}
\label{fig:shock}
\end{figure}

A shock is basically a solution of the hydrodynamics equations
with a discontinuity in the flow of a fluid or plasma.
We will here only consider plane shocks. In the "shock frame" in which
the shock is at rest, the plasma is moving toward the shock with
a velocity $u_1$ from the "upstream" side, and moving away from the
shock on the opposite, "downstream" side with a velocity $u_2<u_1$.
The situation is depicted in Fig.~\ref{fig:shock} where we define
the positive $z-$direction as perpendicular to the shock front pointing
toward the downstream region.
In the Fermi acceleration process, cosmic rays cross back and forth
across astrophysical shocks by scattering on inhomogeneities of the
magnetic fields. As we will show in the following, in each shock crossing,
on average they gain energy, but also have a finite probability for
escaping the acceleration region. This interplay between energy gain
and particle loss will lead to a power law spectrum. We consider here
only the simplest case where the magnetic field is fully turbulent
without any coherent component and we restrict ourselves to the {\it
  test particle regime} in which the back-reaction of accelerated CRs
on the shock properties is neglected. For a comprehensive review of
shock acceleration theory see Ref.~\cite{Blandford:1987pw}.

Let us now consider one cycle of shock crossing of ultra-relativistic
particles, $p\simeq E$, $\beta\simeq1$, for a non-relativistic shock.
In the plasma rest frame where ${\bf E}=0$, the particle momentum is constant.
Let us denote the angle of its direction of motion relative to the $z-$direction
with $\theta_i$, and $\mu_i\equiv\cos\theta_i$, for $i=1,2$. A relativistic
particle of momentum $p_1$ in the upstream rest frame approaching
the shock front has $\mu_1>-u_1$ and will have a momentum $p_2=\Gamma p_1(1+v\mu_1)$
in the downstream rest frame, where $v=(u_1-u_2)/(1-u_1u_2)$ with
corresponding Lorentz factor $\Gamma=(1-v^2)^{-1/2}$. Similarly, a particle
with momentum $p_2$ in the downstream rest frame approaching the shock has
$\mu_2<-u_2$ and will have a momentum
$p_1=\Gamma p_2(1-v\mu_2)$ in the upstream rest frame. After one cycle we
thus have
\begin{equation}\label{eq:one_cycle}
  p_1^\prime=\Gamma^2p_1(1-v\mu_2)(1+v\mu_1)\,.
\end{equation}
Assuming an isotropic distribution in both plasma rest frames and averaging the
fluxes $\propto\mu_i$ over the shock front over
$-u_1\le\mu_1\le1$ and $-1\le\mu_2\le-u_2$, for non-relativistic shocks, $u_1,u_2,v\ll1$
leads to an average momentum gain of
\begin{equation}\label{eq:delta_p}
  \langle\Delta p\rangle\simeq\frac{4}{3}(u_1-u_2)p
\end{equation}
during one cycle, where $p$ is the initial momentum. If there is no
shock, then $u_1-u_2$ and there is no acceleration as it should be
because in this case there is only one reference frame and thus no
electric field in its rest frame that could change the energy of the
particles.

We can estimate the downstream escape probability as the ratio
of the convective flux far from the shock front to the flux
entering the downstream region from the upstream region.
Assuming a constant cosmic ray mass density $n$ downstream, one has
\begin{equation}\label{eq:p_esc}
  P_{\rm esc}\simeq\frac{nu_2}{\frac{n}{4\pi}2\pi\int_0^1d\mu_2\mu_2}=4u_2\,.
\end{equation}
We can now estimate the spectrum as follows: Starting with a
momentum $p_0$, after $n$ cycles, the cosmic rays that have
not escaped the downstream region yet will on average have
momentum $p_n\simeq p_0\left(1+\langle\Delta p\rangle/p\right)^n$,
and a density of $n(>p_n)=n(>p_0)(1-P_{\rm esc})^n$. Writing the
integral spectrum of such cosmic rays as
$n(>p_n)=n(>p_0)(p_n/p_0)^{1-\alpha}$, taking the logarithm yields
\begin{equation}\label{eq:alpha_r}
  \alpha=1-\frac{\ln\left[n(>p_n)/n(>p_0)\right]}{\ln\left(p_n/p_0\right)}
  =\frac{r+2}{r-1}\,,
\end{equation}
where $r\equiv u_1/u_2=\rho_2/\rho_1 > 1$ is the shock compression ratio
because mass conservation implies $\rho_1 u_1=\rho_2 u_2$, where
$\rho_{1,2}$ are the energy densities on the two sides of the shock.
The second equality holds for non-relativistic shocks.

Let us now compute the compression ratio $r$ in terms of the properties of
the plasma. The flow of plasma across the shock front conserves mass, momentum
and energy,
\begin{eqnarray}\label{eq:rh_jump}
  \rho_1u_1&=&\rho_2u_2\,,\nonumber\\
  \rho_1u_1^2+P_1&=&\rho_2u_2^2+P_2\,,\\
  \rho_1u_1\left(\frac{1}{2}u_1^2+h_1\right)&=&
  \rho_2u_2\left(\frac{1}{2}u_2^2+h_2\right)\,,\nonumber
\end{eqnarray}
where $P_i$ and $h_i$ are pressure and specific enthalpy, respectively,
of the plasma on the two sides, $i=1,2$. The specific enthalpy is the
"available" energy of a system under constant pressure. It is given by
$h=(\rho_{\rm kin}+P)/\rho$, where $\rho_{\rm kin}$ is the kinetic (excluding
rest mass) energy density. For an ideal, non-relativistic gas $P\propto\rho$
and thus the speed of sound $c_s^2=dP/d\rho=P/\rho$. Furthermore,
$P/\rho_{\rm kin}=\gamma-1$, where $\gamma$ is the adiabatic index.
For an ideal, non-relativistic gas it is $\gamma=5/3$, so that one has
$P=\frac{2}{3}\rho_{\rm kin}=nT=\frac{\rho}{m}T$, where $m$ and $T$ are
mass and temperature of the particles and we have used
$\rho_{\rm kin}=\frac{1}{2}nm\langle v^2\rangle=\frac{3}{2}nT$ with
$\langle v^2\rangle$ the average squared particle velocity at temperature
$T$. Thus, in this case $c_s^2=T/m$. In general, one thus has
$h=\frac{\gamma}{\gamma-1}c_s^2=\frac{\gamma}{\gamma-1}P/\rho$.

Eqs.~(\ref{eq:rh_jump}) are often called the Rankine-Hugoniot jump conditions.
Dividing the last one by $\rho_1u_1u_2^2=\rho_2u_2^3=\rho_1u_1^3/r^2$ [these
equalities are a consequence of the first condition in Eqs.~(\ref{eq:rh_jump})],
we obtain
\begin{equation}\label{eq:r1}
  r^2\left(\frac{1}{2}+\frac{\gamma}{\gamma-1}\frac{1}{{\cal M}^2}\right)=
  \frac{1}{2}+\frac{\gamma}{\gamma-1}\frac{P_2}{\rho_2u_2^2}=
  \frac{1}{2}+\frac{\gamma}{\gamma-1}r\frac{P_2}{\rho_1u_1^2}\,,
\end{equation}
where we have introduced the upstream Mach number ${\cal M}\equiv u_1/c_{s,1}$.
Using the second condition in Eqs.~(\ref{eq:rh_jump}), we can express
$P_2$ in terms of upstream quantities, $P_2=P_1+\frac{r-1}{r}\rho_1u_1^2$.
Wit this, the right hand side of Eq.~(\ref{eq:r1}) becomes
$-(\gamma+1)/[2(\gamma-1)]+\gamma\left(1+{\cal M}^{-2}\right)r/(\gamma-1)$.
This yields a quadratic equation for $r$ whose largest solution is
\begin{equation}\label{eq:r2}
  r=\frac{\gamma+1}{\gamma-1+2\gamma/{\cal M}^2}\,.
\end{equation}
We now realize that for non-relativistic shocks with $\gamma=5/3$ we have
$r\le4$ and from Eq.~(\ref{eq:alpha_r}) $\alpha\ge2$, where the limiting
values are obtained in the limit of large Mach number, ${\cal M}\gg1$.

We can estimate the time scales associated with acceleration as follows:
In the limit of an infinitely extended shock the escape time is given
by equating the downstream convection and diffusion
distances, $u_2T_{\rm esc}\simeq\left[2D_2(p)T_{\rm esc}\right]^{1/2}$,
with $D_2(p)$ the downstream diffusion coefficient, or
\begin{equation}\label{eq:t_esc}
  T_{\rm esc}(p)\simeq\frac{2D_2(p)}{u_2^2}
\end{equation}
Furthermore, the cycle time scale is $T_{\rm cyc}\simeq P_{\rm esc}T_{\rm esc}$
and the acceleration time scale is $T_{\rm acc}\simeq(p/\langle\Delta p\rangle)T_{\rm cyc}$,
giving
\begin{eqnarray}\label{eq:t_cyc_acc}
  T_{\rm cyc}(p)&\simeq&\frac{D_2(p)}{2u_2}\nonumber\\
  T_{\rm acc}(p)&\simeq&\frac{3D_2(p)}{8(u_1-u_2)u_2}\,,
\end{eqnarray}
where we have used Eqs.~(\ref{eq:p_esc}) and~(\ref{eq:delta_p}).
In cases where collisional energy losses can not be neglected,
one can obtain the maximal energy by comparing $T_{\rm acc}(p)$
to the relevant energy loss time scales, such as inverse Compton
scattering (ICS), synchrotron radiation, pion production, to be discussed
below, as well as with the shock lifetime.

Relativistic shocks with $u_1,v\to1$ are more complicated
to treat because the particle distributions can not be assumed
isotropic anymore. Nevertheless, in the most optimistic scenarios,
one has
\begin{equation}\label{eq:t_acc_rel}
  T_{\rm acc}(p)\sim D(p)\sim r_g(p)\sim\frac{p}{eZB}\,.
\end{equation}
It is obvious from Eq.~(\ref{eq:one_cycle}) that as long as particles
are isotropically distributed in the intervals $-u_1\le\mu_1\le1$
and $-1\le\mu_2\le-u_2$, one has $\left\langle p_1^\prime\right\rangle\sim2\Gamma^2p_1$
for relativistic shocks and thus particles gain energy very
efficiently in one cycle. However, in a highly relativistic shock, the
upstream plasma frame approaches the shock with almost the speed of
light, $u_1\simeq1$, and thus, after returning from the downstream
frame at the end of the first cycle, the CR may not have sufficient
time to isotropize before being caught up by the shock. In this case,
the projection of the CR velocity onto the shock normal $\mu_1$ is
given by the same quantity on the downstream region
by a Lorentz transformations as
$\mu_1\equiv p_{1,z}/p_1=(p_{2,z}-vp_2)/\left[p_2(1-v\mu_2)\right]=
(\mu_2-v)/(1-v\mu_2)$. Therefore, the factor $(1+v\mu_1)$ in
Eq.~(\ref{eq:one_cycle}) becomes $1/[\Gamma^2(1-v\mu_2)]\sim1/(2\Gamma^2)$
since the condition for crossing from downstream to upstream
requires $\mu_2\simeq-1$ for a relativistic shock for which also
$u_2\simeq1$. Using Eq.~(\ref{eq:one_cycle}) again
then yields $\left\langle p_1^\prime\right\rangle\simeq p_1$ and
thus there is little energy gain after the first cycle.

\subsection{Maximal Acceleration Energy}
Let us now estimate the maximal energy up to which cosmic rays can be
accelerated in shocks. In this section we consider all quantities in
the shock rest frame. If the shock itself moves with a Lorentz factor
$\Gamma$, there will be a geometry-dependent relation between
quantities in the shock rest frame and in the observer frame. As an
example for this case we will consider the fireball model for
gamma-ray bursts in more detail in the next section.

At very high energies, Eq.~(\ref{eq:t_esc}) becomes larger than the
diffusion time over the finite linear size $R$ of the shock,
\begin{equation}\label{eq:t_esc2}
  T_{\rm esc}^R\simeq\frac{R^2}{D(p)}\ga\frac{R^2}{r_g(p)}
  \simeq\frac{eZB_{\rm rms}R^2}{p}\,,
\end{equation}
where we have used Eq.~(\ref{eq:bohm}). Since acceleration requires $T_{\rm
  acc}\la T_{\rm esc}^R$, using Eq.~(\ref{eq:t_cyc_acc}) for the acceleration time
scale with $D(p)\ga r_g(p)$, this results in the famous Hillas
criterion~\cite{hillas-araa},
\begin{equation}\label{eq:E_max}
  E_{\rm max}\la eZRBv\simeq10^{18}\,v\,\left(\frac{B}{\mu{\rm G}}\right)
  \left(\frac{R}{\rm kpc}\right)\,{\rm eV}\,,
\end{equation}
where we have abbreviated the root mean square of the magnetic field
$B_{\rm rms}$ by $B$. There is another, quicker and more qualitative
derivation of Eq.~(\ref{eq:E_max}) that also gives an order of magnitude
estimate of the acceleration time scale Eq.~(\ref{eq:t_cyc_acc}): 
The time scale at which diffusion and convection
length scales become comparable in the shock, $d\sim D(p)t\sim(vt)^2$,
is given by $t\sim D(p)/v^2$. Requiring that the corresponding length
scale $d$ is smaller than the shock size, $d\simeq D(p)/v\la
R$ immediately results in Eq.~(\ref{eq:E_max}) when Eq.~(\ref{eq:bohm}), $D(p)\ga
r_g(p)$, with the gyro radius $r_g(p)$ given by Eq.~(\ref{eq:gyro}) is used.

\begin{figure}[ht]
\includegraphics[width=\textwidth,clip=true,angle=0]{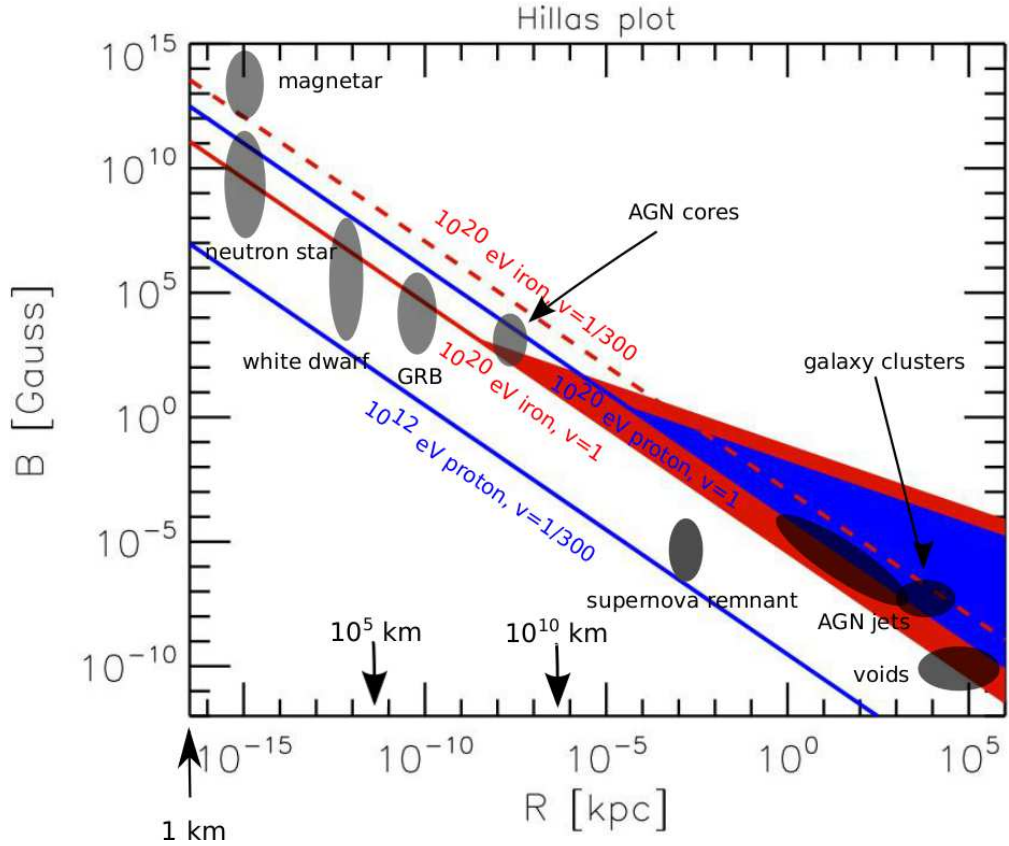}
\caption[...]{The ``Hillas plot'' represents astrophysical objects
  which are potential cosmic-ray accelerators on a two-dimensional
  diagram where on the horizontal direction the size linear
  extension $R$ of the accelerator, and on the vertical
  direction the magnetic field strength $B$ are plotted. According to
  Eq.~(\ref{eq:E_max}), the maximal acceleration energy $E$ is
  proportional to $ZRBv$, where $v$ is the shock velocity in units of the
speed of light and $Z$ is the particle charge. Particular values for the maximal energy correspond
to diagonal lines in this diagram and can be realized either in
a large, low field acceleration region or in a compact accelerator
with high magnetic fields. For a shock velocity $v\sim$1, neutron
stars, AGN, Radio Galaxies or Galactic clusters can accelerate protons
to $E\sim10^{20}\,$eV. For typical non-relativistic shocks,
$v\sim1/300$, as they are realized, for example, in supernova
remnants, no astrophysical objects of sufficient size and magnetic
field to produce $10^{20}\,$eV protons are
known. The blue and red shaded wedges signify the parameter ranges
satisfying both the Hillas condition Eq.~(\ref{eq:E_max}) and the
synchrotron condition Eq.~(\ref{eq:E_max_synch}) for a $10^{20}\,$eV proton and
iron, respectively in the shock rest frame }
\label{fig:hillas}
\end{figure}

Note that for relativistic shocks, $v\simeq1$,
the Hillas criterion is equivalent to the intuitive condition that the
gyro radius has to be smaller than the size of the shock, $r_g(p)\la
R$. The Hillas criterion is shown and compared with various
astrophysical objects in Fig.~\ref{fig:hillas}.

Accelerating particles of charge $eZ$ to an energy $E_{\rm max}$ also
requires a minimal source power which can be estimated as follows:
Acceleration to an energy $E_{\rm max}$ requires an induction ${\cal
  E} \ga E_{\rm max}/(eZ)$. With $Z_0\simeq100\,\Omega$
the vacuum impedance, this leads to the dissipation of a minimal power
of~\cite{Lovelace,Blandford:1999hi}
\begin{equation}\label{eq:Lmin}
  L_{\rm min}\simeq\frac{{\cal E}^2}{Z_0}\simeq 10^{45}\,Z^{-2}\,
  \left(\frac{E_{\rm max}}{10^{20}\,{\rm eV}}\right)^2\,
  {\rm erg}\,{\rm s}^{-1}\,.
\end{equation}
When expressing the square of the product of the magnetic field in an
accelerator with its size in terms of a luminosity, $L\sim B^2R^2$,
this condition can be expressed in terms of the Hillas-criterium
Eq.~(\ref{eq:E_max}) for relativistic shocks, $v\simeq1$, which
states that the gyro radius of a charged particle at the maximal
acceleration energy must fit within the accelerator of size $R$. Eq.~(\ref{eq:Lmin})
suggests that the power requirements are considerably relaxed for
heavier nuclei which is easy to understand because an estimate solely
based on motion of charged particles in magnetic fields can only
depend on their rigidity $E/Z$. 

However, the Hillas criterion Eq.~(\ref{eq:E_max}) and the minimal power
Eq.~(\ref{eq:Lmin}) are necessary but in general not sufficient since
they do not take into
account energy loss processes within the source.
If the approximation of collision-less acceleration is not good, the
maximal energy is further constrained by the condition that the
acceleration time scale must not only be smaller than the escape time,
but also smaller than the energy loss time over which the particle
looses a given fraction of its energy. The most important energy loss
processes for cosmic rays will be discussed in
section~\ref{section4} below. Furthermore, the shock may only
have a finite lifetime $T_{\rm dyn}$ which gives an additional condition. The most
general version of the equation determining the maximal energy is thus
\begin{equation}\label{eq:E_max_loss}
  T_{\rm acc}(p)\la\min\left[T_{\rm esc}^L(p),T_{\rm loss}(p),T_{\rm dyn}\right]\,.
\end{equation}
The modifications resulting form taking into account energy
loss processes have recently been discussed in
Ref.~\cite{Ptitsyna:2008zs}. For example, for diffuse shock
acceleration synchrotron radiation by the accelerated nuclei is the
dominant energy loss mechanism and yields the additional constraint
\begin{equation}\label{eq:E_max_synch}
  E_{\rm max}\la3\times10^{16}\,\frac{A^4}{Z^4}\,\left(\frac{B}{\rm G}\right)^{-2}
  \left(\frac{R}{\rm kpc}\right)^{-1}\,{\rm eV}\,.
\end{equation}
This equation essentially follows from integrating the synchrotron
energy loss rate
\begin{equation}\label{eq:P_syn}
  \left.\frac{dE}{dt}\right|_{\rm syn}\simeq
  -\frac{e^4}{36\pi^2\,m_p^4}\left(\frac{Z}{A}\right)^4\,B^2\,E^2
\end{equation}
over the length scale $R$~\cite{Medvedev:2003sx}, where $m_p$ is the
proton mass. The additional constraint Eq.~(\ref{eq:E_max_synch}) is
shown for $10^{20}\,$eV protons and iron nuclei as the colored wedges
in Fig.~\ref{fig:hillas}.

Acceleration is often assumed to be rigidity limited such
that the differential flux of nuclei of charge $Z$ at energy $E$ can,
for example, be assumed to be of power-law form
\begin{equation}\label{eq:inj_spec}
  \Phi_{Z,A}(E)=\frac{dN_{Z,A}}{dE}\propto q_{Z,A}\,E^{-\alpha}\,\Theta(Z\,E_{\rm max,p}-E)\,,
\end{equation}
where $q_{Z,A}$ is the relative abundance of the nucleus of charge $Z$ and
atomic number $A$ at a given energy $E$, $\alpha$ is the differential
spectral index, $E_{\rm max,p}$ is the maximal proton
energy and $\Theta(x)$ is the Theta-function which cuts off the
spectrum for $E>Z\,E_{\rm max,p}$. To compare with abundances in a non-relativistic
gas, one often refers to abundances $x_{Z,A}$ at a given energy per
nucleon $E/A$, such that for a differential spectrum with power-law
index $s$ one has
\begin{equation}\label{eq:q_x}
  q_{Z,A}\propto x_{A,Z}\,A^{\alpha-1}\,.
\end{equation}
Thus, a steep spectrum can lead to a strong enhancement of the abundance
of heavy nuclei compared to the abundances in the gas at rest.

\subsection{Application to Particle Acceleration in Gamma-Ray Bursts}
In the following we will apply the conditions for the maximal energy
to a particular potential UHECR source class, namely gamma-ray
bursts (GRBs)~\cite{Waxman:1995vg,Dermer:2010km}. In the fireball
model of GRBs an optically thick plasma of photons, leptons and
baryons expands until it becomes transparent to the
$\gamma-$rays. During the ensuing so called ``prompt phase''
the fireball expands with a Lorentz factor $\Gamma\gg1$ which depends on its
baryon content. Since different parts of the fireball will move with
slightly different velocities, different shells will form which will
crash into each other and form shocks at which first order Fermi
acceleration can take place. Finally, when crashing into the
extragalactic medium the fireball will also form
external shocks which would explain the GRB afterglow. The formation
of internal shocks explains the variability on time scales
$t_v\sim0.01\,$s observed during the prompt phase. Since the
individual shells approach the observer with velocities
$v\sim1-1/(2\Gamma^2)$, the typical radius of an internal shock in the
observer frame will be $r_i\simeq t_v/(1-v)\simeq2\Gamma^2 t_v$. Note
that this relation is purely kinematic and not a relativistic effect
because all quantities correspond to the same observer
frame. In the following quantities measured in the comoving rest frame of the
internal shocks will be denoted by a prime and we will in part follow
the notations in Ref.~\cite{Ahlers:2011jj}. We assume
that a fraction $\epsilon_e$ of the turbulent kinetic energy density $u_{\rm
  kin}^\prime$ carried by the baryons, will be converted into
electrons and positrons and a similar part to $\gamma-$rays. The
photon energy density in the shock rest frame is, therefore, given by
$u_\gamma^\prime\simeq\epsilon_e u_{\rm kin}^\prime$. Photon energies
in the observer and the shock rest frame are related by
$\varepsilon=\Gamma\varepsilon^\prime$. Since a length interval $\Delta
x$ measured at equal time in the observer frame, $\Delta t=0$,
transforms as $\Delta x=\Delta x^\prime/\Gamma$, the average photon
number density per volume and energy $n_\gamma(\varepsilon)$ and
$n_\gamma^\prime(\varepsilon^\prime)$ is invariant under Lorentz
transformation and thus one has
$n_\gamma^\prime(\varepsilon^\prime)=n_\gamma(\Gamma\varepsilon^\prime)$. This
implies $u_\gamma^\prime=\int d\varepsilon^\prime\varepsilon^\prime
n_\gamma^\prime(\varepsilon^\prime)=\int d\varepsilon\varepsilon
n_\gamma(\varepsilon)/\Gamma^2\sim L_\gamma/(4\pi\Gamma^2 r_i^2)$, where
$L_\gamma$ is the GRB luminosity in the observer rest frame. For the magnetic
energy density in the shock rest frame $u_B^\prime$ we assume
$u_B^\prime=(B^\prime)^2/(8\pi)\simeq\epsilon_B u_{\rm
  kin}^\prime$. Observations indicate
$\epsilon_B\simeq\epsilon_e\simeq0.1$. After elimination of $u_{\rm
  kin}^\prime$, one can estimate the magnetic field in the shock rest
frame as
\begin{equation}\label{eq:Bprime}
  B^\prime\simeq \left(\frac{\epsilon_e}{\epsilon_B}\right)^{1/2}\,
  \frac{L_\gamma/2}{\Gamma^3\,t_v}\simeq5\times10^4\,
  \left(\frac{\epsilon_e}{\epsilon_B}\right)^{1/2}\,
  \left(\frac {L_\gamma}{10^{52}\,{\rm erg}\,{\rm
        s}^{-1}}\right)^{1/2}\,
  \left(\frac {300}{\Gamma}\right)^3\,\left(\frac {0.01\,{\rm
        s}}{t_v}\right)
  \,{\rm G}\,.
\end{equation}
This also allows to understand the rough structure of the GRB
$\gamma-$ray spectrum: In the shock frame the characteristic
energy of the electrons and positrons will be of the order $\epsilon_e
m_p$. These electrons and positrons will emit synchrotron photons in
the magnetic field Eq.~(\ref{eq:Bprime}) with a characteristic energy in the
observer frame of
\begin{equation}\label{eq:epsilon_c}
  \varepsilon_0\simeq\Gamma\frac{3eB^\prime}{2m_e^3}\,(\epsilon_e
  m_p)^2\simeq0.88\,\epsilon_e^{3/2}\epsilon_B^{1/2}\left(\frac {L_\gamma}{10^{52}\,{\rm erg}\,{\rm
        s}^{-1}}\right)^{1/2}\,\left(\frac {300}{\Gamma}\right)^3\,
  \left(\frac {0.01\,{\rm s}}{t_v}\right)\,{\rm MeV}\,,
\end{equation}
with $m_e$ the electron mass. The observed $\gamma-$ray spectrum
$n_\gamma(\varepsilon)$ of GRBs can be approximated by a broken power
law with an index $-\alpha\simeq-1$
for $\varepsilon\la\varepsilon_0$ and index $-\beta\simeq-2.2$
for $\varepsilon\ga\varepsilon_0$, where the observed break energy
$\varepsilon_0\sim1\,$MeV, consistent with Eq.~(\ref{eq:epsilon_c}).

The acceleration time scale in the shock frame Eq.~(\ref{eq:t_acc_rel})
for relativistic shocks then becomes
\begin{eqnarray}\label{eq:t_acc_GRB}
  T_{\rm acc}^\prime(E)&\sim&\frac{\eta E^\prime}{eB^\prime}\sim\left(\frac{2\epsilon_e}{\epsilon_B
      L_\gamma}\right)^{1/2}\,\frac{\eta\Gamma^2E\,t_v}{e}\nonumber\\
  &\simeq&6.6\times10^{10}\,\left(\frac{\epsilon_e}{\epsilon_B}\right)^{1/2}\,
   \left(\frac{10^{52}\,{\rm erg}\,{\rm s}^{-1}}{L_\gamma}\right)^{1/2}\,\eta\,
    \left(\frac{\Gamma}{300}\right)^2\,\left(\frac{E}{300\,{\rm
          EeV}}\right)\,\left(\frac{t_v}{0.01\,{\rm s}}\right)\,{\rm cm}\,.
\end{eqnarray}
We have put $Z=1$ here because nuclei are quickly photo-disintegrated
in GRBs~\cite{Anchordoqui:2007tn}.

One condition on the maximal acceleration energy is
now obtained by comparing Eq.~(\ref{eq:t_acc_GRB}) with the dynamical
and energy loss times scales, according to the condition
Eq.~(\ref{eq:E_max_loss}). The dynamical times scale in the shock frame where
$\Delta x^\prime$ is given by
\begin{equation}\label{eq:t_dyn_GRB}
T_{\rm dyn}^\prime\simeq\frac{r_i}{\Gamma}\simeq2\Gamma
t_v\simeq1.8\times10^{11}\,\left(\frac{\Gamma}{300}\right)\,\left(\frac{t_v}{0.01\,{\rm
      s}}\right)\,{\rm cm}\,,
\end{equation}
which, together with the characteristic magnetic field strength in the
shock rest frame Eq.~(\ref{eq:Bprime}) is also marked ``GRB'' in
Fig.~\ref{fig:hillas}. Note that because of relativistic boosting, the
energy which has to be reached in the shock rest frame is smaller than the
observed energy $E$ by a factor $\Gamma$. Altogether,
\begin{equation}\label{eq:E_max1_GRB}
  E_{\rm max}\la8.2\times10^{20}\,\left(\frac{\epsilon_B}{\epsilon_e}\right)^{1/2}\,
   \left(\frac {L_\gamma}{10^{52}\,{\rm erg}\,{\rm s}^{-1}}\right)^{1/2}\,
   \left(\frac{1}{\eta}\right)\,\left(\frac {300}{\Gamma}\right)\,{\rm eV}\,,
\end{equation}
which is independent of $t_v$. In fact, this is a generalization of
Eq.~(\ref{eq:Lmin}) to which it reduces for $\Gamma=1$. We now
consider the constraint on the maximal energy from synchrotron
radiation by the protons. The synchrotron cooling time scale is
\begin{eqnarray}\label{eq:t_synch_GRB}
  T_{\rm synch}^\prime(E)&\sim&\frac{(6\pi)^2m^4}{e^4E^\prime
    (B^\prime)^2}=\frac{2(6\pi)^2m^4}{e^4}\,\left(\frac{\epsilon_e}{\epsilon_B
      L_\gamma}\right)\,\frac{\Gamma^7\,t_v^2}{E}\nonumber\\
  &\simeq&8.4\times10^{12}\,\left(\frac{\epsilon_e}{\epsilon_B}\right)\,
   \left(\frac{10^{52}\,{\rm erg}\,{\rm s}^{-1}}{L_\gamma}\right)\,
    \left(\frac{\Gamma}{300}\right)^7\left(\frac{300\,{\rm
          EeV}}{E}\right)\,\left(\frac{t_v}{0.01\,{\rm s}}\right)^2\,{\rm cm}\,.
\end{eqnarray}
Comparing this with Eq.~(\ref{eq:t_acc_GRB}) leads to
\begin{equation}\label{eq:E_max2_GRB}
  E_{\rm max}\la3.4\times10^{21}\,\left(\frac{\epsilon_e}{\epsilon_B}\right)^{1/4}\,
   \left(\frac{10^{52}\,{\rm erg}\,{\rm s}^{-1}}{L_\gamma}\right)^{1/4}\,\frac{1}{\eta^{1/2}}\,
    \left(\frac{\Gamma}{300}\right)^{5/2}\,\left(\frac{t_v}{0.01\,{\rm s}}\right)^{1/2}\,{\rm eV}\,.
\end{equation}

Finally, we estimate photo-hadronic energy loss rates, namely pion
production of the accelerated protons on the radiation field
within the GRB, $p\gamma\to N\pi$. In the shock frame pion production
of a proton of energy $E^\prime$ can only occur with photons of energy
$\varepsilon^\prime\ga\varepsilon_{N\pi}^\prime\sim m_\pi m_p/(2
E^\prime)$, with $m_\pi$ the pion mass, as
can be seen from Eq.~(\ref{eq:gzk}) below. Far above the threshold the
cross section can be approximated by $\sigma_{N\pi}\simeq150\,\mu$b
and the protons loose a substantial fraction of their energy in each
interaction. Therefore, the pion production loss time can be estimated
from
\begin{equation}\label{eq:t_pion_GRB}
  \frac{1}{T_{N\pi}^\prime(E^\prime)}\sim\sigma_{N\pi}\varepsilon_{N\pi}^\prime
  n_\gamma^\prime(\varepsilon_{N\pi}^\prime)\simeq\frac{\sigma_{N\pi}m_\pi
    m_p}{2E^\prime}\,n_\gamma\left(\frac{\Gamma m_\pi m_p}{2E^\prime}\right)\,.
\end{equation}
Since $n_\gamma(\varepsilon)\propto\varepsilon^{-1}$ for
$\varepsilon\la\varepsilon_0$, one has $\varepsilon
n_\gamma(\varepsilon)\sim L_\gamma/(4\pi\varepsilon_0r_i^2)$ for
$\varepsilon\la\varepsilon_0$. Therefore, the interaction rate
Eq.~(\ref{eq:t_pion_GRB}) is roughly energy independent above the
energy
\begin{equation}\label{eq:E_pi_GRB}
  E_{p\pi}\simeq\frac{\Gamma^2m_\pi m_p}{2\varepsilon_0}\simeq
  4\times10^{16}\,\left(\frac{\Gamma}{300}\right)^2\,
  \left(\frac{{\rm MeV}}{\varepsilon_0}\right)\,{\rm eV}
\end{equation}
and corresponds to an energy loss time of
\begin{eqnarray}\label{eq:t_synch_GRB2}
  T_{N\pi}^\prime(E>E_{p\pi})&\sim&\frac{16\pi\varepsilon_0\,\Gamma^5\,t_v^2}
  {\sigma_{N\pi}\,L_\gamma}\nonumber\\
  &\simeq&3.5\times10^{11}\,\left(\frac {\varepsilon_0}{{\rm MeV}}\right)
  \,\left(\frac{10^{52}\,{\rm erg}\,{\rm s}^{-1}}{L_\gamma}\right)
  \,\left(\frac{\Gamma}{300}\right)^5\,\left(\frac{t_v}{0.01\,{\rm
        s}}\right)^2\,
  {\rm cm}\,.
\end{eqnarray}
Comparing this with Eq.~(\ref{eq:t_acc_GRB}) finally gives the additional
condition
\begin{equation}\label{eq:E_max3_GRB}
  E_{\rm max}\la1.6\times10^{21}\,\left(\frac {\varepsilon_0}{{\rm
        MeV}}\right)\,
  \left(\frac{\epsilon_B}{\epsilon_e}\right)^{1/2}\,
   \left(\frac{10^{52}\,{\rm erg}\,{\rm s}^{-1}}{L_\gamma}\right)^{1/2}\,\frac{1}{\eta}\,
    \left(\frac{\Gamma}{300}\right)^3\,\left(\frac{t_v}{0.01\,{\rm s}}\right)\,{\rm eV}\,.
\end{equation}

Although GRBs individually have
more than adequate power to achieve the required maximal acceleration
energies, they are disfavored in terms of the local power density they
can provide in the form of UHECRs~\cite{Eichler:2010ky} compared
to an UHECR origin in AGNs and radio galaxies.

\section{Anisotropies and Nature of the Sources}

Although the main goal of the IceCube experiment~\cite{icecube} originally was the detection
of extraterrestrial high energy neutrinos via up-going muons, it is also an
excellent detector of galactic CRs via the down going muon tracks (see
section~\ref{section6} below) that are produced by the interactions of
cosmic rays in the Earth's atmosphere. By May
2010 it collected about $3\times10^{10}$ such muons with
a median energy of $\simeq20\,$TeV
which provides sufficient statistics to be sensitive to CR anisotropies down to
a level of $\sim10^{-4}$ at degree scales and
above~\cite{Toscano:2011dc}. This revealed anisotropies at a level of
$\simeq10^{-3}$ on scales between $10^\circ$ and $30^\circ$ on the
Northern hemisphere. Anisotropies on a similar level and with a
consistent structure were also
observed by the Tibet AS$\gamma$ array, the Super-Kamiokande Detector,
the Milagro Gamma Ray Observatory, and the ARGO-YBJ and EAS-TOP
experiments (see Ref.~\cite{Toscano:2011dc} for
references). Anisotropies at a similar level, but with a different
structure on the sky was also observed up to energies of
$\simeq400\,$TeV~\cite{Aglietta:2009,Abbasi:2011zk}. This can be explained by
diffusion in the turbulent Galactic magnetic field.

No significant anisotropies have been observed between
$\simeq10^{15}\,$eV and up to the onset of the GZK threshold energy
around a few times $10^{19}\,$eV (see section~\ref{section4} below)
above which UHECRs loose energy within about 20 Mpc and sources
must be located roughly within this distance scale.

In the UHECR regime, the Pierre Auger
Observatory which observes the Southern hemisphere from Argentina
has accumulated enough statistics to detect first signs
of anisotropy: A significant correlation with the 12th edition of the
V\'eron-Cetty and V\'eron catalog of nearby AGNs was observed for
events with energies above $56\,$EeV~\cite{auger-anisotropy}. This is
very suggestive because this is essentially the same threshold energy
above which the GZK effect limits the range of primary cosmic rays to $\sim50\,$Mpc.
This does not necessarily mean that the V\'eron-Cetty and V\'eron
catalog AGNs are the long sought UHECR sources,
but it suggests that the real UHECR sources follow an anisotropic distribution
similar to the one of nearby AGNs. Perhaps not surprisingly the
sources may be astrophysical accelerators which follow the local large scale
structure. Unfortunately, with accumulation of more data, these correlations have
weakened somewhat~\cite{arXiv:1009.1855}. The fraction of events above
55 EeV correlating with the Veron Cetty Catalog has came down from
$69^{+11}_{-13}$\% to $38^{+7}_{-6}$\% compared to 21\% expected
for isotropy, while at the same time the statistical significance has
stayed roughly constant at $\simeq99\,$\% confidence level. If one
divides the sky distribution into an isotropic component and a component
correlating, for example, with the 2MASS redshift survey, this corresponds to a relatively large
isotropic fraction of 60--90\%~\cite{arXiv:1009.1855}. There is also a
significant excess correlations seen with the 2MASS redshift survey.
On the other hand, observations by the HiRes and Telescope Array
experiments in the Northern
hemisphere~\cite{Abbasi:2008md,Thomson:2010tc} are so far consistent
both with isotropy and with the fraction of events observed by the
Pierre Auger Observatory that correlate with the Veron Cetty Catalog.

An interesting argument linking UHECR sources to their luminosity at radio
frequencies has been put forward in this context by
Hardcastle~\cite{Hardcastle:2010cq}. He concludes that if UHECRs are
predominantly protons, then very few sources should
contribute to the observed flux. These sources should be easy to
identify in the radio and their UHECR spectrum should cut off steeply
at the observed highest energies. In contrast, if the composition is
heavy at the highest energies then many radio galaxies could
contribute to the UHECR flux but due to the much stronger deflection
only the nearby radio galaxy Centaurus A (``Cen A'') may be identifiable.

The Pierre Auger data does indeed show a clustering of
super-GZK events towards the direction of Centaurus
A (NGC 5128)~\cite{arXiv:1009.1855}. This is somewhat surprising since,
although Cen A is the closest radio galaxy at a distance of just
$\simeq3.6\,$Mpc and the third-strongest
radio source in the sky, it is an elliptical
radio galaxy with a relatively small power
output~\cite{Rieger:2009pm} which makes it difficult to reach the
required UHECR energies. On the other hand, UHECR events observed
towards Cen A could originate mainly from sources within the
Centaurus galaxy cluster which is itself part of the Hydra-Centaurus
supercluster and located just behind Cen A. In
any case, Cen A has been observed in many
channels and its small distance allows detailed astronomical and
astrophysical studies. For example, its lobes have been detected in 200 MeV
gamma-rays by Fermi Large Area Telescope (Fermi LAT)~\cite{cenA-lobes-fermi}, and its core was
observed by Fermi LAT~\cite{cenA-core-fermi}. Such observations and
its potential role as a major local UHECR accelerator has
triggered many multi-messenger model building efforts for Cen
A~\cite{Rieger:2009pm,cenA-tomas}.
For example, it has been pointed out in Ref.~\cite{cenA-tomas} that proton acceleration
in the jet of Cen A is hard to reconcile with observations of Cen A in
TeV gamma-rays by the H.E.S.S. telescopes~\cite{cenA-hess} if gamma-rays are produced
by proton-proton interactions. Instead, p$-\gamma$ interactions in the core are
more consistent with these observations. We note that no significant
excesses have been observed toward other directions on the sky where one could expect an
overdensity of potential UHECR accelerators. For example, the Virgo
cluster at a distance of $\simeq20\,$Mpc, thus within the GZK horizon,
contains the prominent radio galaxy M87 but shows no excess UHECR
flux~\cite{Gorbunov:2007ja}. 

\section{Propagation and Deflection}
\label{section4}
\subsection{Galactic Cosmic Rays}
Galactic cosmic rays at least up to energies around the knee are deep
in the diffusive regime in which their propagation can be described by
solving a diffusion-convection-energy loss equation for the
location- and energy dependent CR density per energy interval
$n_i=n_i({\bf r},p)$ of a nuclear species $i$ which depends on location
and energy. It has the following general form,
\begin{eqnarray}\label{eq:galactic_propagation}
\partial_t n_i&=&\partial_a\left(D_{ab}\partial_b-{\bf v_c}_a\right)n_i+
\partial_p\left(p^2D_{pp}\partial_p\frac{n_i}{p^2}\right)-
\partial_p\left[\dot p\,n_i-\frac{p}{3}\left(\nab\cdot {\bf
      v_c}n_i\right)\right]+\nonumber\\
&&+\sum_{j>i}\beta n_{\rm gas}\int dp^\prime\sigma_{j\to
  i}(p,p^\prime)n_j(p^\prime)-\beta
\sigma_i(p)n_{\rm gas}+\Phi_i\,,
\end{eqnarray}
where $\beta$ is the CR velocity, $D_{ab}$ is the diffusion tensor, ${\bf v}_c({\bf r})$
is the convection velocity of the galactic plasma, $D_{pp}({\bf r},p)$ describes
diffusion in momentum space and can give rise to re-acceleration,
$\dot p(p)$ describes energy loss processes, $\sigma_{j\to
  i}(p,p^\prime)$ is the cross section for producing species $i$ by
spallation of species $j$ upon interaction with the gas
of density $n_{\rm gas}({\bf r})$, $\sigma_i(p)$ is the total inelastic cross
section for interactions with the gas and $\Phi_i({\bf r},p)$ is the CR
injection rate density per energy
interval. To avoid clutter, in Eq.~(\ref{eq:galactic_propagation}) we
have only kept the dependencies on ${\bf r}$ and $p$ which are
essential for an unambiguous definition of the various terms.

In general, Eq.~(\ref{eq:galactic_propagation}) has
to be solved numerically within a given geometry for the galactic disc
which often is approximated as a slab with cylindrical symmetry such
that all quantities only depend on $r$ and $z$. In
addition, assuming isotropic spatial
diffusion with $D_{ab}({\bf r},p)=\delta_{ab}D({\bf r},p)$ in a
turbulent magnetic field the scalar diffusion coefficient can be
parametrized as
\begin{equation}\label{eq:diff_coeff}
  D({\bf r},p)=D_0\beta\left(\frac{p/Z}{{\rm GV}}\right)^\delta\,f(r,z)\,,
\end{equation}
with the rigidity $p/Z$ in units of GigaVolts and where
$f(r,z)$ is often taken as $f(r,z)\propto\exp(|z|/z_0)$, with $z_0$
the disc scale height. Under these assumptions and within quasilinear
theory the momentum diffusion coefficient is related to $D$
by~\cite{seo-ptuskin}
\begin{equation}\label{eq:Dpp_Dxx}
  D_{pp}(p)=\frac{4v_A^2p^2}{3\delta(4-\delta^2)(4-\delta)D(p)}\,.
\end{equation}
There are several public program packages available such as
GalProp~\cite{galprop} and DRAGON~\cite{dragon} for solving
Eq.~(\ref{eq:galactic_propagation}) in this scenario. The
parameters $\delta$ and $D_0/z_0$ can be fixed by measuring secondary
to primary CR ratios such as the born to carbon or the nitrogen to
oxygen ratio~\cite{DiBernardo:2009ku}. Typical values are
$\delta\simeq$0.4--0.5 and $D_0/z_0\simeq\mbox{(0.6--1)}\times10^{28}\,{\rm
  cm}^2\,{\rm s}^{-1}\,{\rm kpc}^{-1}$.

With this we can now make a
very simple argument linking the Galactic CR spectrum injected at the
sources and the one observed at Earth: The supernova remnants that are
thought to accelerate Galactic cosmic rays are now routinely observed
in $\gamma-$rays and typically have a spectrum $\propto E^{-\alpha}$
with $\alpha\simeq2.2$. It is also thought that at least for some of these
supernova remnants there is a significant, if not dominant hadronic
contribution to the $\gamma-$ray emission in which primary protons and
nuclei interact with the ambient
gas~\cite{Morlino:2011di,Aharonian:2011da}. Since the cross section for this interaction
depends on energy only logarithmically, the slopes of the observed
$\gamma-$ray spectrum and of the primary cosmic ray spectrum emitted
by the sources should be essentially equal. An acceleration spectrum
$\propto E^{-\alpha}$ with $\alpha\simeq2.2$ is also consistent with
non-relativistic shock acceleration theory, as we have seen in
chapter~\ref{section2}. Assuming simple diffusion
within a ``leaky box'' of height $\simeq2z_0$ with a diffusion
coefficient $D(p)\propto p^\delta$ that is spatially constant within
the box, the {\it confinement time} in the
Galactic disk can be estimated as $T_{\rm conf}(p)\propto
z_0^2/D(p)\propto p^{-\delta}$. Therefore, the charged cosmic ray
spectrum observed at Earth $n(p)$ and the injected spectrum per volume
$\Phi(p)\propto p^{-\alpha}$ are related by
\begin{equation}\label{eq:spec_inj_obs}
  n(p)\sim\Phi(p)\,T_{\rm conf}(p)\propto\frac{\Phi(p)}{D(p)}\propto
    p^{-\alpha-\delta}\,.
\end{equation}
For $\alpha\simeq2.2$, as observed through the $\gamma-$rays, and
$\delta\simeq0.5$ for the power law index of the energy dependence of
the diffusion coefficient, as inferred from Galactic CR nuclei
abundances, this gives $n(p)\propto p^{-2.7}$ which is roughly
consistent with observations, see Fig.~\ref{fig:cr-spectrum}.

\subsection{Extra-Galactic Cosmic Rays}
We first discuss the main interaction processes of extragalactic
hadronic UHECR here; the interactions of secondary $\gamma-$rays and
neutrinos are the subject of section~\ref{section7}. Besides redshift which leads to a continuous
decrease of the absolute momentum, $dp/dt=H(z)p$, with $H(z)$ the
Hubble constant at a given redshift $z$, hadronic UHECR can interact
with non-relativistic baryons in the form of gas or with low energy
target photons. UHECR interactions with Galactic and extragalactic gas
correspond to a cross section of order 0.1 barn, leading to an
interaction time scale of order $\sim10^{14}\,(n_g/10^{-6}\,{\rm
  cm}^{-3})\,yr$, with $n_g$ the gas density. In general this has a
negligible influence on UHECR propagation, except possibly in the
centers of galaxy clusters or when secondary $\gamma-$rays and
neutrinos from hadronic interactions are considered. Interactions with
low energy photons includes Compton scattering, which is also
negligible, pair production, also known as Bethe-Heitler process, pion
production and, for nuclei, photo-disintegration. In the following we
consider the latter three processes.

The interaction length $l(E)$
of a CR of energy $E$ and mass $m$ propagating through 
a background of target photons is given by
\begin{equation}
  l(E)^{-1}=\int d\varepsilon n_b(\varepsilon)\int_{-1}^{+1}d\mu
  \frac{1-\mu\beta}{2}\,\sigma(s)=\frac{1}{2\gamma^2}\int d\varepsilon
  \frac{n_b(\varepsilon)}{\varepsilon^2}\int_0^{2\gamma\varepsilon}
  d\varepsilon^\prime\varepsilon^\prime\sigma(\varepsilon^\prime)
  \,,\label{eq:intlength}
\end{equation}
where $n_b(\varepsilon)$ is the number density of the target photons
per unit energy at energy $\varepsilon$,  
$\beta=(1-m^2/E^2)^{1/2}$ is the CR velocity, $\mu$ is the cosine
of the angle between the incoming momenta, $\gamma=E/m$ is the CR
Lorentz factors and $\sigma(s)$ is the total cross section of the relevant process
for the squared center of mass (CM) energy
\begin{equation}
  s=m_b^2+m^2+2\varepsilon E\left(1-\mu\beta\right)=2m\varepsilon^\prime
  \,.\label{eq:s}
\end{equation}
The last expressions in Eqs.~(\ref{eq:intlength}) and~(\ref{eq:s})
result from a Lorentz transformation to the CR rest frame in which the
photon has energy $\varepsilon^\prime$. This form is particularly
useful when using data from laboratory measurements which are usually
performed in the rest frame of the nucleus, such that the cross
sections $\sigma(s)$ are usually known as functions of
$\varepsilon^\prime$.
The most relevant target photon backgrounds turn out to be the
infrared, the CMB, and the radio background. A review of
the universal photon background has been given in Ref.~\cite{rt}.

The effective energy loss rate $dE/dt$ is given by multiplying the integrand
in Eq.~(\ref{eq:intlength}) by $E\eta(s)$ where the inelasticity $\eta(s)$, i.e. the
fraction of the energy transferred from the incoming
CR to the recoiling final state particle of interest, is given by
\begin{equation}
  \eta(s)\equiv1-\frac{1}{\sigma(s)}\int dE^\prime
  E^\prime\frac{d\sigma}{dE^\prime}\,
  (E^\prime,s)\,,\label{eq:etas}
\end{equation}
where $E^\prime$ is the energy of the recoiling
particle considered in units of the incoming CR energy
$E$. Here by recoiling particle we usually mean the ``leading''
particle, i.e. the one which carries most of the energy.

Pair production by a nucleus $^A_Z$ with mass number $A$ and charge
$Z$ on a photon $\gamma$, $^A_Z\gamma\to^A_Z e^+e^-$, has a threshold
energy of
\begin{equation}
  E_{\rm npp}=\frac{m_e(m+m_e)}{\varepsilon}\simeq
  4.8\times10^{17}\,A\,\left(\frac{\varepsilon}{10^{-3}\,{\rm
        eV}}\right)^{-1}\,{\rm eV}\,,\label{eq:thr_npp}
\end{equation}
where $\varepsilon\sim10^{-3}\,$eV
represents the energy of a typical target photon such as a CMB photon.
The inelasticity is very small, $\eta\sim10^{-3}$, such that pair
production can be treated as a continuous energy loss process with
energy loss rate~\cite{PSB}
\begin{equation}
  \left.\frac{dE_{A,Z}}{dt}\right|_{\rm NPP}=-\frac{3\alpha
    Z^2\sigma_T(m_e T_b)^2}{(2\pi)^3}f\left(\frac{m_e}{2\gamma T_b}\right)
    \,.\label{eq:rate_npp}
\end{equation}
Here, we have assumed a thermal background of temperature $T_b$ because
the CMB is by far the dominating background, $\alpha\simeq1/137$ is
the fine structure constant, $\sigma_T\equiv e^4/(6\pi m_e^2)$ is the
Thomson cross section, and $f(x)$ is a function that was parametrized
in Refs.~\cite{Blumenthal:1970nn,chodorowski}.

Let us next discuss pion production by UHE nucleons. The threshold
for the reaction $N\gamma\to N\pi$, for
a head-on collision of a nucleon $N$ of energy $E$ and mass $m_N$ with
a photon of energy $\varepsilon$ is given by
\begin{equation}
  E_{N\pi}=\frac{m_\pi(m_N+m_\pi/2)}{2\varepsilon}\simeq
  3.4\times10^{19}\left(\frac{\varepsilon}{10^{-3}\,{\rm eV}}\right)^{-1}
  \,{\rm eV}\,,\label{eq:gzk}
\end{equation}
The pion production cross section that enters Eq.~(\ref{eq:intlength})
and~(\ref{eq:etas}) has a threshold $\varepsilon^\prime\simeq150\,$MeV
in the nucleon rest frame, has a resonance at
$\varepsilon^\prime\simeq350\,$MeV with a peak value of
$\sigma_{N\pi}(350\,{\rm MeV})\simeq600\,\mu$b and an
asymptotically almost constant cross section
$\sigma_{N\pi}(\varepsilon^\prime\ga{\rm GeV})\simeq150\,\mu$b in
the multi-pion production regime. The inelasticity is $\simeq20$\%
close to the threshold and $\simeq50$\% far above the threshold. Also,
there is a channel which
conserves the charge of the original nucleon, and one in which a
protons turns into a neutron and a neutron turns into a proton. The
charge conserving channel produces mostly neutral pions which decay
into secondary $\gamma-$rays, whereas charge exchange reactions
produce mostly charged pions which eventually decay into secondary
electrons or positrons and neutrinos. These are the main production
channels of secondary photons and neutrinos by hadronic cosmic rays,
either within the sources, or during propagation to the observer, as
we will see in more detail in section~\ref{section7}. Pion production
is often modelled with the help of the numerical package SOPHIA~\cite{sophia}.

To good approximation pion production by nuclei can be described in the
{\it superposition model} in which nuclei are treated as a
superposition of $Z$ free protons and $A-Z$ free neutrons and their
binding energy is neglected. Then the cross section for the reaction
$^A_Z\gamma\to ^A_Z\pi$ is given by
\begin{equation}
  \sigma_{^A_Z\pi}(E)=Z\sigma_{p\pi}(E/A)+(A-Z)\sigma_{n\pi}(E/A)
  \,.\label{eq:sigma_pion}
\end{equation}
Note that the energy carried away by a pion in such an
interaction is $\sim20$\% of the interaction nucleon energy, hence
only $\sim20$\%/$A$ of the primary nucleus. The threshold for
photo-pion production is also increased to $\simeq 4\,10^{19}\times
A$eV.

The range of energies for the photodisintegration process, in terms of
the photon energy $\varepsilon^\prime$ in the rest frame of the nucleus,
splits into two parts. The first contribution comes from the low
energy range up to 30 MeV, in the Giant Dipole Resonance region, where
emission of one or two nucleons dominates; the second contribution
comes from energies between 30 MeV and 150 MeV, where multi-nucleon
energy losses are involved. These processes have first been discussed
in some detail in Ref.~\cite{PSB}. More recent descriptions can be
found in Refs.~\cite{Khan:2004nd}. In a photodisintegration event the
changes in energy, $\Delta E$, and
atomic number, $\Delta A$, are related by $\Delta E/E=\Delta A/A$.
Thus, the effective energy loss rate due to photodisintegration
$\left.\frac{dE}{dt}\right|_{\rm eff,photo}$ is given by
\begin{equation}
\frac{1}{E}\left.\frac{dE}{dt}\right|_{\rm
  eff,photo}=\frac{1}{A}\frac{dA}{dt}=\sum_i\,\frac{i}{A}\,l_{A,i}(E)\,,
\end{equation}
where $l_{A,i}(E)$ is the mean free path Eq.~(\ref{eq:intlength}) for emission of
$i$ nucleons.

\begin{figure}[h!]
\includegraphics[width=0.7\textwidth]{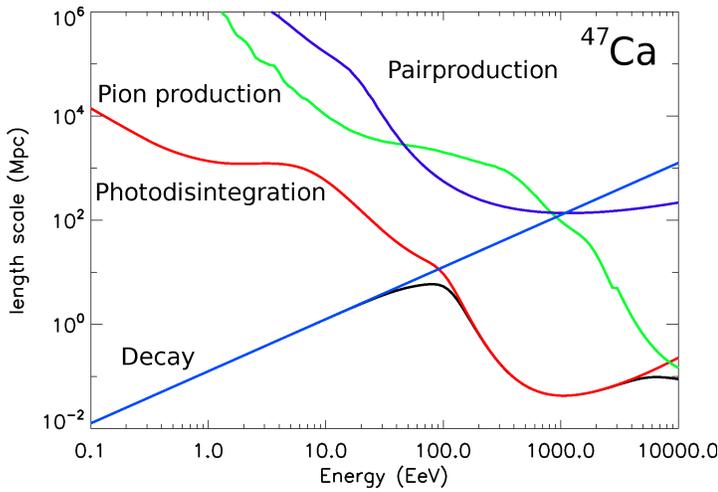}
\caption{ The length scales for all relevant processes of the example
  nucleus $^{47}$Ca as a function of energy: Energy loss length for pair production, decay
  length, and mean free path for pion production and
  photodisintegration.}
\label{fig:Ca47_lengths}
\end{figure}

Fig.~\ref{fig:Ca47_lengths} shows the lengths scales for all relevant interactions for
the example nucleus $^{47}$Ca.

The amount of deflection of extragalactic cosmic rays in cosmic
magnetic fields is still hard to quantify. In a
field with rms strength $B$ and coherence length $l_c$ the rms
deflection angle of a cosmic ray of energy $E$ and charge $Ze$
traveling a distance $d$ is given by~\cite{Waxman:1996zn}
\begin{eqnarray}\label{eq:deflec}
  \theta(E,d)&\simeq&\frac{(2dl_c/9)^{1/2}}{r_g}\\
  &\simeq&0.8^\circ\,
  Z\left(\frac{E}{10^{20}\,{\rm eV}}\right)^{-1}
  \left(\frac{d}{10\,{\rm Mpc}}\right)^{1/2}
  \left(\frac{l_c}{1\,{\rm Mpc}}\right)^{1/2}
  \left(\frac{B}{10^{-9}\,{\rm G}}\right)\,,\nonumber
\end{eqnarray}
in the small deflection angle limit $\theta(E,d) \ll1$, where
$r_g$ is again the gyro radius of Eq.~(\ref{eq:gyro}). The dependence on the
dimension-full quantities can be easily understood from the fact that
the deflection angles $\theta_1\simeq l_c/r_g$ accumulated in
individual domains of size $l_c$ in which the magnetic field can be
approximated as being constant, are uncorrelated between different
domains and thus have to be added in quadrature.

For an order of magnitude estimate for
the deflection angles in the Galactic magnetic field we use $l_c\sim100\,$pc,
$d\sim10\,$kpc, $B\sim3\,\mu$G gives
$\theta(E)\sim1^\circ\,Z(10^{20}\,{\rm eV}/E)$. Thus, protons around
the GZK cut-off, $E\sim60\,$EeV, will be deflected by a few degrees or
less, whereas iron nuclei can be deflected by several tens of
degrees. This implies that Galactic
magnetic fields are likely to destroy any possible correlation with the
local large scale structure in case of a heavy composition.
Numerical simulations using detailed models of the Galactic magnetic
field, including a turbulent component, demonstrate that the relatively large
deflections of UHECR nuclei in the Galactic magnetic field can
considerably distort the images of individual sources. If UHECR source
follow the local large scale structure its image can also be distorted
and shifted considerably~\cite{Giacinti}.

Deflection of UHECRs in cosmic magnetic fields goes along with a time delay
compared to the light travel time, which can be quantified by~\cite{Waxman:1996zn}
\begin{equation}
  \tau(E,d)\simeq d\theta(E,d)^2/4\simeq1.5\times10^3\,Z^2
  \left(\frac{E}{10^{20}\,{\rm eV}}\right)^{-2}
  \left(\frac{d}{10\,{\rm Mpc}}\right)^{2}
  \left(\frac{l_c}{1\,{\rm Mpc}}\right)
  \left(\frac{B}{10^{-9}\,{\rm G}}\right)^2\,{\rm yr}\,,
  \label{eq:delay}
\end{equation}
which, up to numerical factors, follows from elementary geometry of an
arc of length $d$ curving with an angle $\theta(E,d)$. This has
interesting consequences for UHECR sources that are variable
or bursting on time scales comparable or smaller than $\tau(E,d)$:
Provided that the time scale over which the UHECR sky is observed
(typically over a few years) is smaller than $\tau(E,d)$, the
observable spectrum from such an intermittent source will be peaked
around a specific energy, independent of the actual source spectrum:
Higher-energy particles would have passed the observer already,
whereas lower-energy particles would not have arrived yet. This could
happen in particular for GRB sources.

Large scale extra-galactic magnetic fields (EGMF) are much less well
known than Galactic magnetic fields~\cite{EGMF-reviews}. One of
the major detection methods for the EGMF is to measure the Faraday rotation of the
polarization of radio emission from a distant source which is
proportional to the line of sight integral of the product of the plasma density and the
parallel magnetic field component. These so called Faraday rotation
measures are only sensitive to magnetic fields stronger than
$\sim0.1\mu$G, whereas fields down to $\sim$nano Gauss still have
significant effects on UHECR deflection, according to
Eq.~(\ref{eq:deflec}). To measure Faraday rotations in such weaker fields
requires much higher statistics than is currently available. The
statistical average over the sky of an all pervading EGMF is
currently constrained to be $\la3\times10^{-7}\,(l_c/{\rm
  Mpc})^{1/2}\,$G~\cite{Blasi:1999hu}.
Assuming an EGMF whose flux is frozen into the highly conducting
plasma of the intergalactic medium and thus follows the large scale
structure gives the more stringent limit $B\la10^{-9}-10^{-8}\,$G for
an all pervading field. At the same time, the fields in the sheets and
filaments of the galaxy distribution can in this case be as strong as a
micro Gauss. Fields of that strength are also routinely observed in
galaxy clusters which are the largest virialized structures in the
Universe. Outside of galaxy clusters there are currently at most hints
of an EGMF, for example
in the Hercules and Perseus-Pisces superclusters~\cite{Xu:2005rb}.
Within the next 10-15 years, large scale radio
telescopes such as Lofar and SKA are, however, expected to
dramatically improve observational
information on the EGMF in the large scale structure.

From the theoretical point of view, the EGMF in the voids are expected to be very
weak and uncontaminated by astrophysical processes. This makes voids
excellent probes of relic magnetic seed fields that may have been
created in the early Universe~\cite{Grasso:2000wj}. It is very
interesting in this context that the non-observation at GeV
energies by the Fermi LAT satellite
experiment of certain distant blazars that have been detected at TeV
energies by the ground based H.E.S.S. experiment suggest a {\it lower limit} $B\ga3\times10^{-16}\,$G
on the EGMF in the voids~\cite{Neronov:2010gk}. This is because the TeV gamma-rays
observed by H.E.S.S. would initiate electromagnetic cascades that should be
detectable by Fermi unless an EGMF of strength
$B\ga3\times10^{-16}\,$G deflects the electrons and positrons in these
cascades into a diffuse halo around the source whose flux is then
below the Fermi LAT sensitivity. We note in this context, however, that magnetic fields
$\la10^{-12}\,$G are not relevant for UHECR propagation.

\begin{figure}[ht!]
\includegraphics[width=0.7\textwidth]{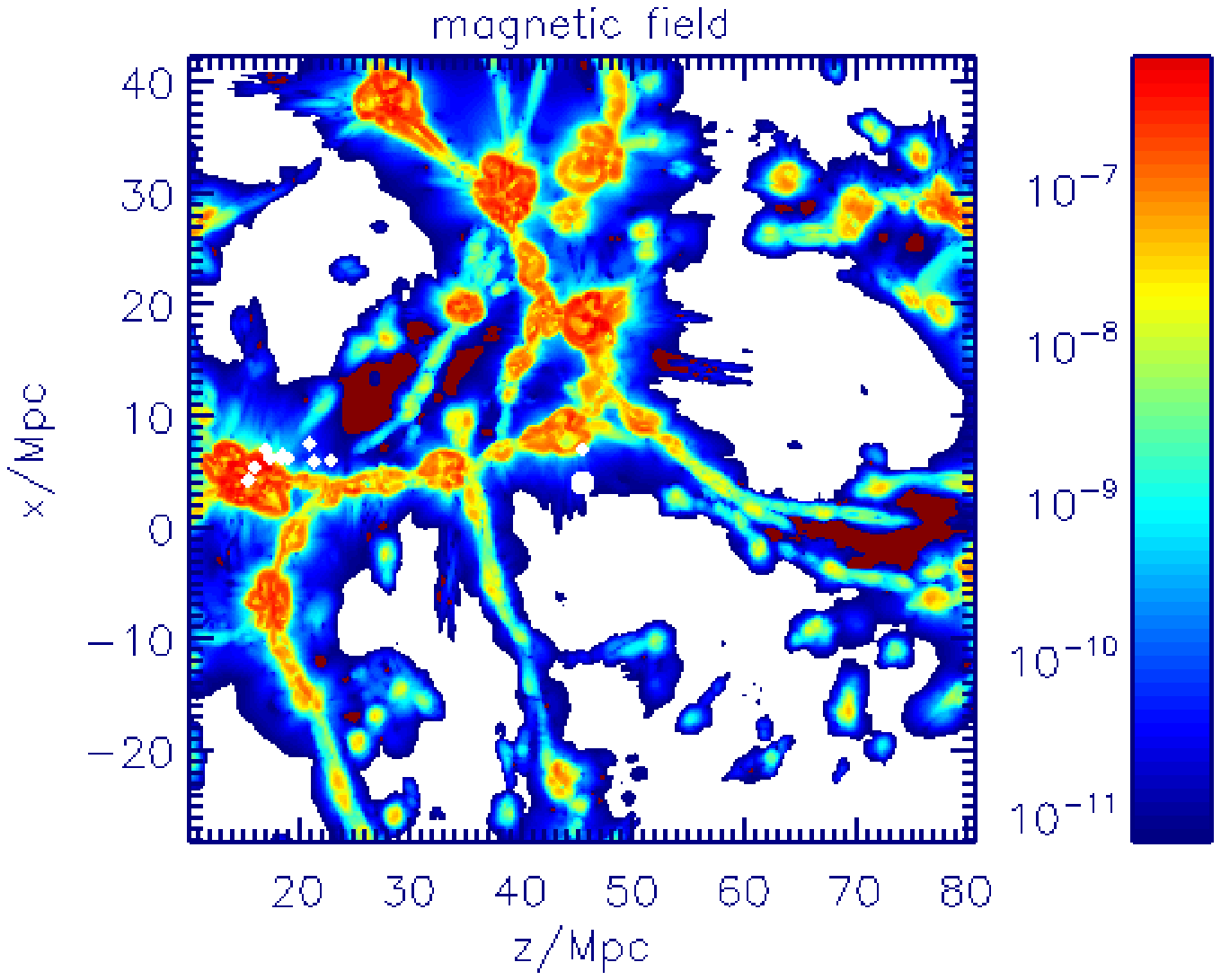}
\includegraphics[width=0.7\textwidth]{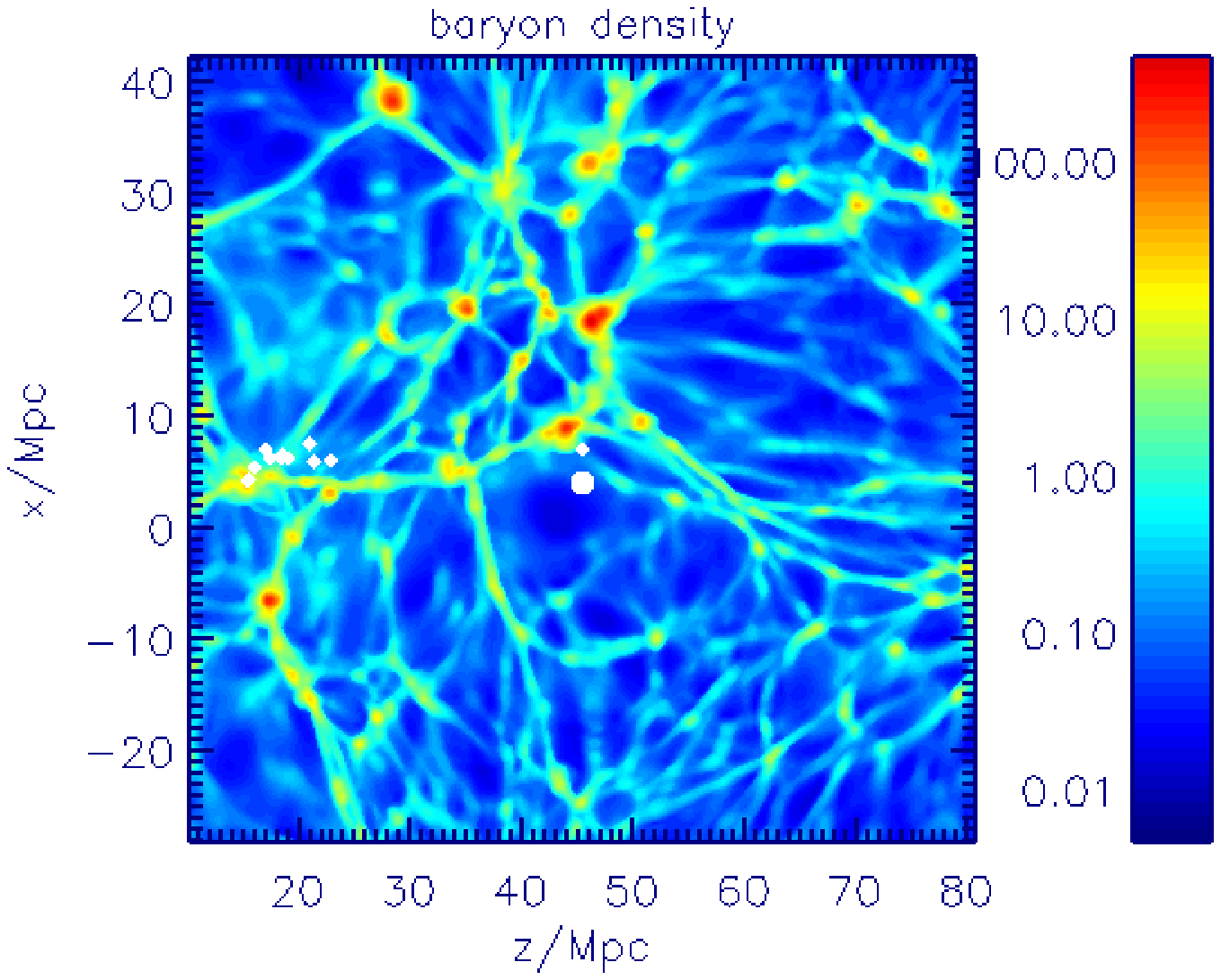}
%
%
\caption{A cross section through the large scale structure simulation
from Ref.~\cite{ryu,miniati} on a scale of 70 Mpc in both
directions. Ten sources are marked with white diamonds in the environment of a massive galaxy cluster. The
white sphere of radius 1 Mpc indicates the observer and right above
marked as a further white diamond is a nearby source. Upper panel:
Magnetic field strength as color contours in units of Gauss, as
indicated. Lower panel: Baryon density is color contours in units of
average density.}
\label{fig:LSS1}       
\end{figure}

\begin{figure}[h!]
\includegraphics[width=\textwidth]{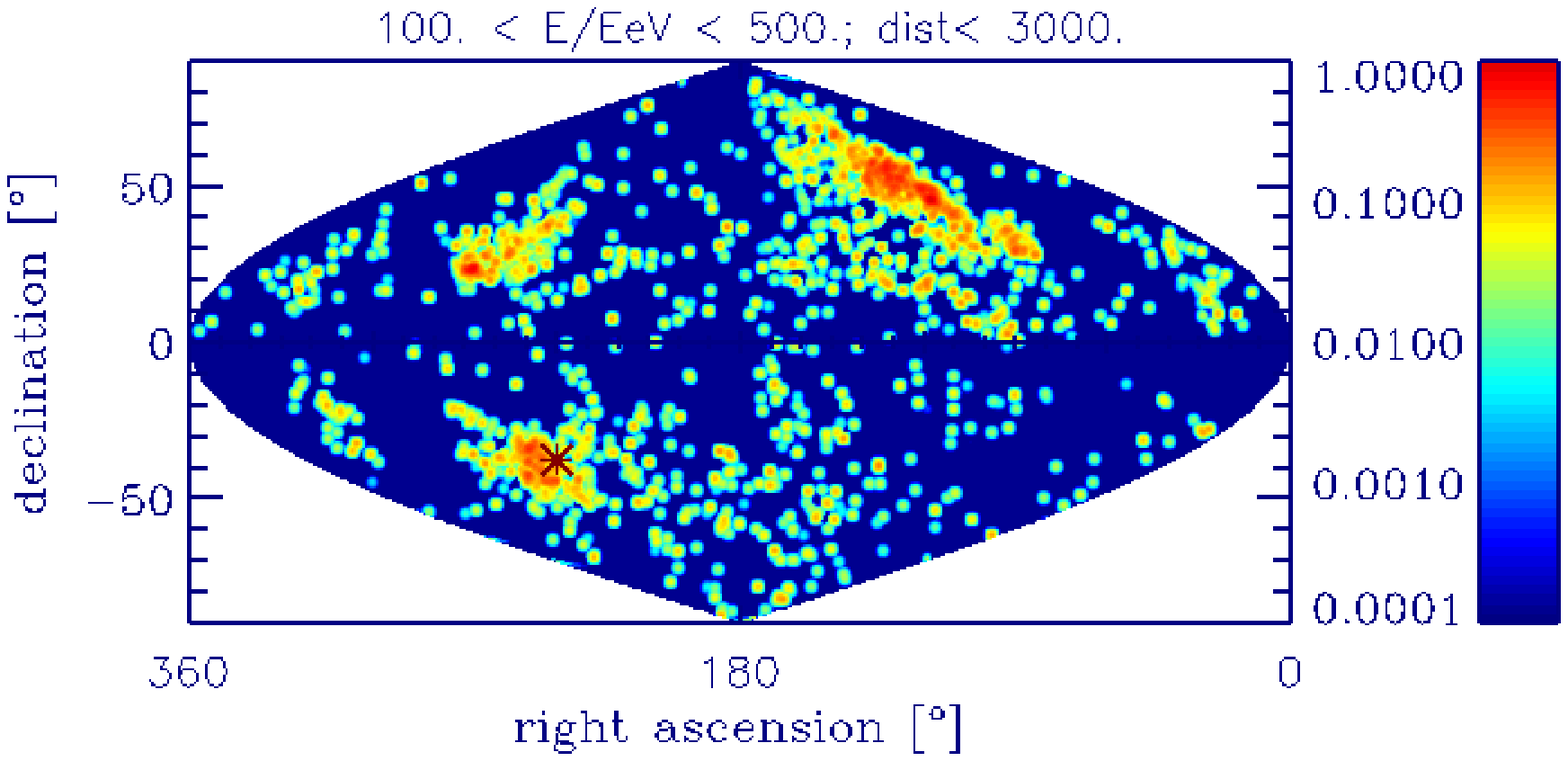}
%
%
\caption{Simulated arrival directions of UHECR above $10^{20}\,$eV in
a scenario where the sources shown in Fig.~\ref{fig:LSS1} inject
mostly protons with a small admixture of a fraction of
$x_{26,56}=6.83\times10^{-4}$ iron nuclei at given energy per nucleon, $E/A$,
roughly corresponding to the galactic abundance of iron. Acceleration
is taken to be rigidity limited power law
$\Phi_{Z,A}(E)=dN_{Z,A}/dE\propto q_{Z,A}\,E^{-2.7}\Theta(Z\,E_{\rm max,p}-E)$
with $E_{\rm max,p}=3.85\times10^{20}\,$eV, where according to
Eq.~(\ref{eq:q_x}) the iron abundance at given energy $E$ is
$q_{26,56}=0.64$. All sources are assumed to have equal
luminosity. The star marks the location of the prominent galaxy
cluster in the middle left part of the cross sections in Fig.~\ref{fig:LSS1}}
\label{fig:UHECR_sky}       
\end{figure}

While waiting for more observational information on the EGMF, another
approach for quantifying the effects of large scale cosmic magnetic
fields on UHECR propagation is to build models of the EGMF
using large scale structure simulations. Two major techniques used in
this context are a magnetohydrodynamic (MHD) version of a constrained smooth particle
hydrodynamics code~\cite{EGMF-Dolag} and Eulerian grid-based hydro+n-body
codes~\cite{EGMF-Miniati}. The magnetic fields are followed by solving
the MHD equations in the ambient plasma while neglecting back-reaction
onto the plasma. Since the MHD equations are linear in the magnetic
field, a seed field is required which is assumed to be either uniform
or concentrated around cosmic shocks, for example at accretion shocks
around galaxy clusters, where they could arise astrophysically, for example,
through the Biermann battery mechanism. The magnetic field is then
normalized such that it approximately reproduces the largest fields observed in galaxy
clusters. Alternatively, it has been assumed that the EGMF follows the
local vorticity and turbulent energy density of the intergalactic plasma~\cite{EGMF-Ryu}.
All these numerical approaches agree on the fact that these fields
tend to follow the large scale galaxy distribution such that they tend to be strongest
around the largest matter concentrations. A cross section through one of
these simulations~\cite{ryu,miniati} is shown in Fig.~\ref{fig:LSS1} (upper
panel). On the other hand, numerical models disagree on certain aspects that are
relevant for UHECR deflection, most notably the distribution of filling factors, i.e.
the fraction of space filled with EGMF above a certain strength, as a function of
that strength~\cite{Sigl:2004gi}. While this can cause considerable differences in the 
size of the deflection angles predicted between the source and the
observed events, before impinging on the Galactic magnetic field, the
UHECR arrival directions should still follow the large scale galaxy
distribution because the deflections tend to be
{\it along and within} the cosmic large scale structure.
This is demonstrated in Fig.~\ref{fig:UHECR_sky} where the deflected
UHECR arrival directions tend to follow arc-like structures that
result from deflections within the large scale cosmic filaments.
In other words, the EGMF is unlikely to deflect UHECRs out of the
large scale structure since the fields in the voids
are very small. Furthermore, the local group of galaxies is a
relatively inactive region and not strongly magnetized such that
extragalactic deflection within $\simeq$1--2 Mpc is likely to be
small. This means that while the events will in general not point back to their
sources, the arrival direction distribution of UHECRs
coming from outside the Galaxy is likely to still correlate
with the local large scale structure even in scenarios with strong
EGMF, heavy nuclei and large deflection angles. Therefore, for
a heavy composition the main effect on correlations with the local
large scale structure will come from the in this case substantial deflections in the
Galactic magnetic fields which should not correlate with extragalactic deflections.

Structured EGMF can also have considerable effects on spectrum and
composition of the cosmic ray spectrum at energies $\sim10^{18}\,$eV
and below, in particular for sources within strongly magnetized
structures: The magnetic fields in galaxy clusters reach micro-Gauss
strength over the core regions which can stretch over several 100 kpc,
with coherence lengths of 10--100 kpc. Eq.~(\ref{eq:delay}) then implies that
UHECR of energy $E\la\hbox{few}\times\,Z\times10^{17}\,$eV will remain
confined during the lifetime $\la10^{10}\,$yr of the galaxy
cluster. This can have two effects: First, below these energies the spectrum
from such sources will be strongly suppressed, which amounts to a
``magnetic horizon
effect''~\cite{Stanev:2000fb,Globus:2007bi}. Second, the mass
composition at a given energy in this range can be strongly shifted to
a light composition compared to the injected composition because
heavier ions are confined longer and thus more heavily suppressed.

The simulations for Fig.~\ref{fig:UHECR_sky} have been performed with
the public software package CRPropa~\cite{Armengaud:2006fx,crpropa}
which can propagate of UHE nuclei with and without deflection, including
secondary electromagnetic cascades and neutrinos.

\section{Mass Composition}
\begin{figure}[ht]
\includegraphics[width=0.7\textwidth,clip=true,angle=0]{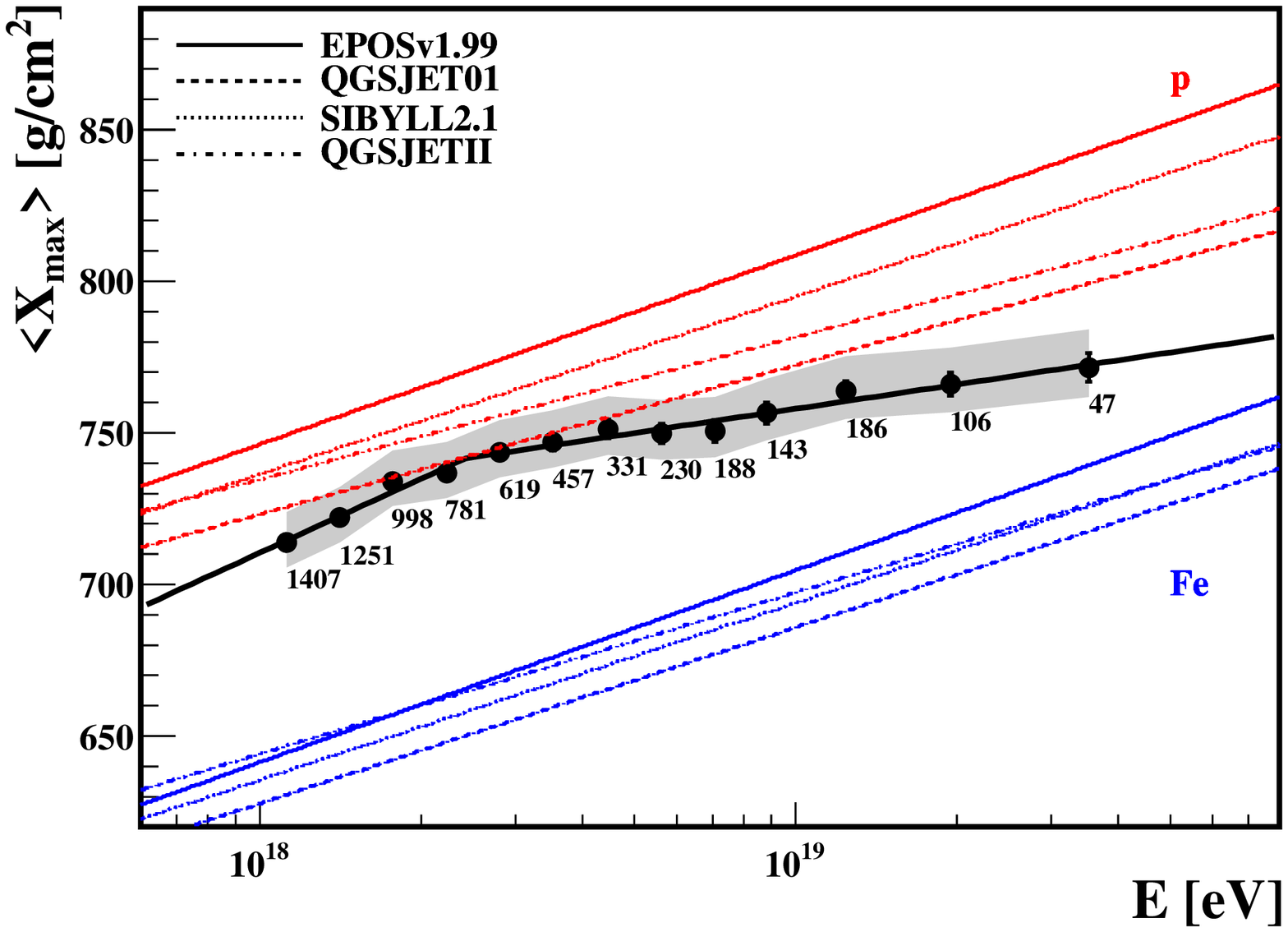}
\includegraphics[width=0.7\textwidth,clip=true,angle=0]{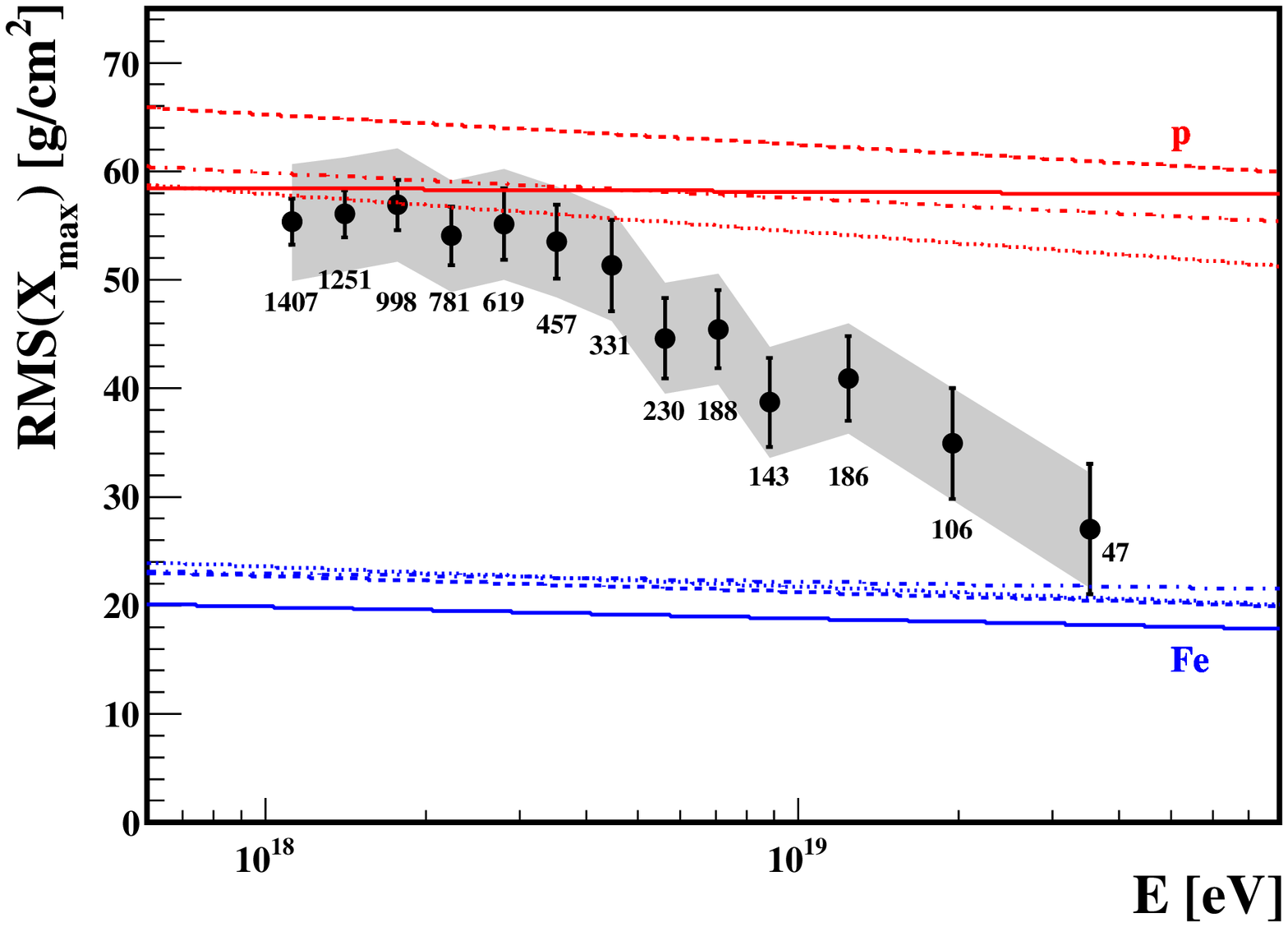}
\caption[...]{Air shower data from the Pierre Auger Observatory: The
  average atmospheric depth at which the showers in a
  given energy bin peak (upper panel) and its root mean square (lower
  panel), as a function of energy, compared to the predictions of
  different air shower simulations for proton (red, upper curves) and
  iron primaries (blue, lower curves). Plots are taken from Ref.~\cite{auger:2011pe}.}
\label{fig:auger-composition}
\end{figure}
Another interesting new question concerns the mass composition
of highest energy cosmic rays: The depth in the atmosphere where
particle density in the giant air showers observed by the Pierre Auger
Observatory is maximal, and in particular the fluctuations of the
depth of shower maximum from event to event, when compared with air
shower simulations, point towards a composition that gradually becomes
heavier with increasing energies~\cite{Abraham:2010yv,auger:2011pe},
as is evident in Fig.~\ref{fig:auger-composition}. It is also interesting to
note in this respect that there is a tension between the average
atmospheric depth and its fluctuation at the highest energies, when
interpreted within the hadronic interaction models shown in
Fig.~\ref{fig:auger-composition}: Whereas the
average depth hints at a mixed composition with a significant light
component, the observed fluctuations become so
small that they would be more consistent with an almost pure iron
composition.

On the other hand, HiRes observations are consistent with a light
composition above $\simeq1.6\times10^{18}\,$eV and up to
$\simeq5\times10^{19}\,$eV above which statistics is insufficient to
determine composition~\cite{Abbasi:2009nf}. This could indicate that statistics is
still too limited to draw firm conclusions or that the Northern and
Southern hemispheres are significantly different in terms of UHECR
composition. In addition, there are
significant uncertainties in hadronic cross sections, multiplicities and
inelasticities and none of the existing hadronic interaction models
consistently describes the shower depth and muon data of the Pierre Auger
experiment~\cite{Ulrich:2009hm,auger:2011pe2}. Note that the center of mass energy
for a UHECR interacting in the atmosphere reaches a PeV$=10^{15}\,$eV,
which is still a factor of a few hundred higher than the highest
energies reached in the laboratory, at the Large Hadron Collider (LHC)
at CERN. It is, therefore, not excluded that the true
mass composition is light on both hemispheres and the UHECR data teaches
us something about hadronic interactions at energies unattainable in
the laboratory.

The question of mass composition is linked to other observables
such as the UHECR spectrum. Unfortunately, the current statistics is
still insufficient to gain significant information on the mass
composition from the observed spectrum. The flux suppression observed
above $\simeq4\times10^{19}\,$eV is qualitatively consistent
with either proton or nuclei heavier than carbon up to iron
nuclei~\cite{Allard:2008gj,Anchordoqui:2007tn,Anchordoqui:2007fi}. In
the latter case, the main energy loss process responsible for the
``cut-off'' is photo-disintegration on the CMB and infrared
backgrounds. It should be noted, however, that the observed flux
suppression could also be due to the intrinsic maximal acceleration
energies attained in the sources, although it would possibly be
somewhat of a coincidence that this energy should be close to the GZK
energy.

\section{High Energy Neutrino Detection}
\label{section6}
Detection rates of high energy neutrinos will not only depend on their
fluxes for which we have discussed various scenarios in the two
previous two chapters, but also on their interaction cross section
with the detector medium. We, therefore, will now discuss the
scattering cross sections of neutrinos with ordinary matter.

Imagine a neutrino of energy $E_\nu$ scattering on a parton $i$
carrying a fraction $x$ of the 4-momentum $P$ of a state $X$ of mass
$M$. Denoting the fractional recoil energy of $X$ by
$y\equiv E_X^\prime/E_\nu$ and the distribution of parton type $i$ by
$f_i(x,Q)$, in the relativistic limit $E_\nu\gg m_X$ the contribution to
the $\nu X$ cross section turns out to be
\begin{equation}
  \frac{d\sigma_{\nu X}}{dxdy}=\frac{2G_{\rm F}^2ME_\nu x}{\pi}
  \left(\frac{M_{W,Z}^2}{2ME_\nu xy+M_{W,Z}^2}\right)^2
  \sum_i f_i(x,Q)\left[g_{i,L}^2+g_{i,R}^2(1-y)^2\right]\,.\label{nuX}
\end{equation}
Here, $g_{i,L}$ and $g_{i,R}$ are the left- and right-chiral couplings
of parton $i$, respectively. Eq.~(\ref{nuX})
applies to both charged and neutral currents, as well as to the
case where $X$ represents an elementary particle such as the
electron, in which case $f_i(x,Q)=\delta(x-1)$.

As usual, if the four-momentum transfer $Q$ becomes comparable to
the electroweak scale, $Q^2\gg M_{W,Z}^2$, the weak gauge boson
propagator effects, represented by the factor
$M_{W,Z}^2/(Q^2+M_{W,Z}^2)$ in Eq.~(\ref{nuX}), become important.
We have used that in the limit $|Q^2|\gg M^2$ one has
$0\simeq-M^2=(xP+Q)^2\simeq Q^2+2P\cdot Qx$ with
$P\cdot Q\simeq-ME_X^\prime=-ME_\nu y$ evaluated in the laboratory
frame, i.e. the rest frame of $X$ before the interaction.
$Q^2\simeq2ME_\nu xy$ is also called the {\it virtuality}
because it is a measure for how far the exchanged gauge boson
is form the mass shell $Q^2=-M_{W,Z}^2$.

We will not derive Eq.~(\ref{nuX}) in detail, but it is easy to
understand its structure: First, the overall normalization is
analogous to the cross section in the four-fermion approximation,
\begin{equation}
  \sigma(\bar{\nu}_ep\to ne^+)=\frac{G_{\rm F}^2}{\pi}|M_{if}|^2
  \frac{p_f^2}{v_iv_f}\,,\label{cross2}
\end{equation}
using the fact that for $E_\nu\gg M$
the CM momentum $p_*^2\simeq M E_\nu/2$. Second, if the helicities of
the parton and the neutrino are equal, the total spin is zero
and the scattering is spherically symmetric in the CM frame.
In contrast, if the parton is right-handed, the total spin is
1 which introduces an angular dependence: After a rotation by
the scattering angle $\theta_*$ in the CM frame the particle
helicities are unchanged for the outgoing final state particle
and one has to project back onto the original helicities in order
to conserve spin. If a left-handed particle originally
propagated along the positive z-axis, its left-handed component
after scattering by $\theta_*$ in the $x-z$ plane is
\begin{equation}
  \frac{1}{2}\left(1-\frac{\sss\cdot{\bf p}}{p}\right)
  \left(\matrix{0\cr1}\right)
  =\frac{1}{2}\left(\matrix{-\sin\theta_*\cr1+\cos\theta_*}\right)\,,
\end{equation}
giving a projection $[(1+\cos\theta_*)/2]^2$. Now, Lorentz transformation
from the CM frame to the lab frame gives $E_\nu^\prime/E_\nu=
(1+v_X\cos\theta_*)/2\simeq(1+\cos\theta_*)/2$ in the relativistic
limit and thus the projection factor equals $(E_\nu^\prime/E_\nu)^2=
(1-E_X^\prime/E_\nu)^2=(1-y)^2$, as in Eq.~(\ref{nuX}) for the
right-handed parton contribution. Integrated over $0\leq y\leq1$
this gives $1/3$, corresponding to the fact that only one of
the three projections of the $J=1$ state contributes.

We now briefly consider neutrino-nucleon interaction.
From Eq.~(\ref{nuX}) it is obvious that for $2E_\nu m_N\ll M_{W,Z}^2$
the total neutrino-nucleon cross section is proportional to the
neutrino energy $E_\nu$. In the opposite, ultra-high energy limit
$2E_\nu m_N\gg M_{W,Z}^2$, the dominant contribution comes
from partons with
\begin{equation}
  x\sim\frac{M_{W,Z}^2}{2E_\nu m_N}\,.
\end{equation}
Since, very roughly, $xf_i(x,Q)\propto x^{-0.3}$ for $x\ll1$, it follows
that the neutrino-nucleon cross section approximately grows $\propto E_\nu^{0.3}$.
This is confirmed by a more detailed evaluation of Eq.~(\ref{nuX})
using the CTEQ4--DIS parton distributions~\cite{Gandhi:1998ri}. A good
power law fit is given by
\begin{equation}
  \sigma_{\nu N}(E)\simeq2.36\times10^{-32}(E/10^{19}
  \,{\rm eV})^{0.363}\,{\rm cm}^2\quad(10^{16}\,{\rm eV}\la
  E\la10^{21}\,{\rm eV})\,.\label{eq:sigma_nuN}
\end{equation}

Let us use this to do a very rough estimate of event rates
expected for extraterrestrial UHE neutrinos in neutrino telescopes.
Such neutrinos are usually produced via pion production by
accelerated UHE protons interacting within their source or
with the cosmic microwave background (CMB) during propagation to
Earth. At GZK energies the secondary neutrino flux should very roughly
be comparable with the primary UHE cosmic ray flux, within large
margins. This argument is actually known as ``Waxman-Bahcall bound'',
see Eq.~(\ref{eq:wb}) below. Using that the
neutrino-nucleon cross section from Eq.~(\ref{eq:sigma_nuN}) scales as
$\sigma_{\nu N}\propto E_\nu^{0.363}$ for
$10^{16}\,{\rm eV}\la E_\nu\la10^{21}\,$eV, and assuming
water or ice as detector medium, we obtain the rate
\begin{eqnarray}
  \Gamma_\nu&\sim&
  \sigma_{\nu N}(E_\nu)2\pi E_\nu j(E_\nu)n_N V_{\rm eff}\nonumber\\
  &\sim&0.03\left(\frac{E_\nu}{10^{19}\,{\rm eV}}\right)^{-0.637}
  \left(\frac{E_\nu^2 j(E_\nu)}
  {10^2\,{\rm eV}{\rm cm}^{-2}{\rm sr}^{-1}{\rm s}^{-1}}\right)
  \left(\frac{V_{\rm eff}}{{\rm km}^3}\right)\,{\rm yr}^{-1}\,,\label{nurate}
\end{eqnarray}
where $n_N\simeq6\times10^{23}\,{\rm cm}^{-3}$ is the nucleon
density in water/ice, $V_{\rm eff}$ the effective detection volume,
and $j(E_\nu)$ is the differential neutrino flux in units of
${\rm cm}^{-2}{\rm eV}^{-1}{\rm sr}^{-1}{\rm s}^{-1}$.

Eq.~(\ref{nurate}) indicates that at $E_\nu\ga10^{18}\,$eV,
effective volumes $\ga100\,{\rm km}^3$ are necessary. Although
impractical for conventional neutrino telescopes, big air
shower arrays such as the Pierre Auger experiment~\cite{Abraham:2009uy} can achieve
this. In contrast, if there are sources such as active galactic
nuclei emitting at $E_\nu\sim10^{16}\,$eV at a level
$E_\nu^2 j(E_\nu)\sim10^2\,{\rm eV}{\rm cm}^{-2}{\rm sr}^{-1}{\rm s}^{-1}$,
km-scale neutrino telescopes should detect something.

\begin{figure}[ht]
\includegraphics[width=0.9\textwidth]{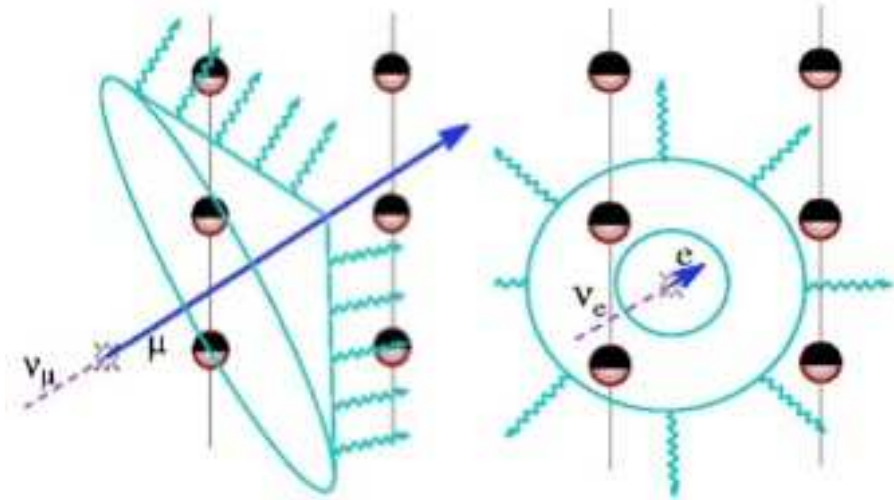}
\caption[...]{Left panel: Muon neutrino detection in water or ice by
  Cherenkov radiation of the muon produced by a charged current
  interactions. Right panel: Neutrino detection by electromagnetic and
hadronic cascades.}
\label{fig:nu_detection}
\end{figure}

Nowadays there are several different techniques to detect high energy
neutrinos~\cite{nu_review}. The most conventional detects the Cherenkov emission of the
high energy muons in water or ice that are produced by charged current
interactions of muon neutrinos. Interactions which do not produce
muons, namely neutral current interactions and charged current
interactions of electron and tau-neutrinos, can also be detected via
the electromagnetic and hadronic cascades initiated by the electrons
and tau-leptons and by the energy-momentum transferred to the target
nuclei. These two detection modes are sketched in
Fig.~\ref{fig:nu_detection}. At the time of writing of this monograph,
active experiments using this technique include the Lake Baikal neutrino
telescope~\cite{lake-baikal}, the ANTARES
neutrino telescope~\cite{antares} which operates off the coast of Toulon in the
Mediterranean sea and the IceCube telescope~\cite{icecube} in the ice beneath the
South Pole, which is currently is the largest instrument and will
reach a sensitivity of $E_\nu^2j(E_\nu)\simeq2\,{\rm eV}\,{\rm cm}^{-2}\,{\rm
    sr}^{-1}\,{\rm s}^{-1}$ to the diffuse neutrino flux between
  $\simeq10^{14}\,$eV and $\simeq2\times10^{16}\,$eV after 3 years
  of operation with 86 strings carrying a total of 5,160 optical
  modules. An infill is planned to lower the threshold down to about
  10 GeV. Due to the large background of down-going atmospheric muons
  from cosmic ray interactions in the atmosphere above the detector,
  these instruments are usually optimized to detection of up-going
  neutrinos, which limits their energy range to $\la10^{16}\,$eV above
  which neutrinos get absorbed in the Earth. However, the atmospheric
  neutrino background becomes sufficiently small again above
  $\simeq10^{17}\,$eV to detect down-going neutrinos. To complement
  IceCube which operates in the Southern hemisphere, there are plans
  to construct a km-scale neutrino telescope in the Northern
  hemisphere, based on the three smaller scale and water-based
  projects ANTARES, NEMO and NESTOR.

\begin{figure}[ht]
\includegraphics[width=0.9\textwidth]{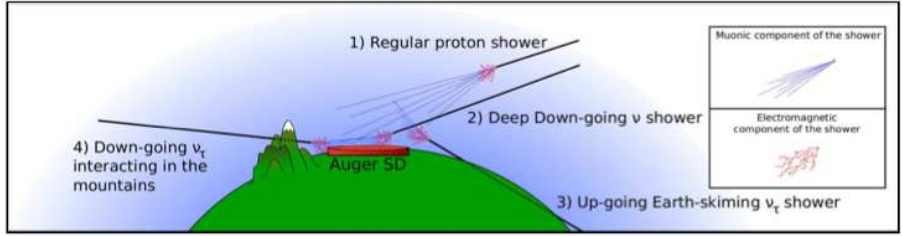}
\caption[...]{Sketch of the neutrino detection modes of the Pierre
  Auger Observatory (from Ref.~\cite{auger:2011pf}).}
\label{fig:nu_auger}
\end{figure}

Neutrinos with energies above $\sim10^{17}\,$eV can induce giant air
showers in the atmosphere, just as charged cosmic rays and
$\gamma-$rays do. However, due to their small cross section, the
neutrino interaction probability in one vertical atmospheric depth is
$\la10^{-4}$ even at the highest energies. Therefore, neutrino induced
air showers are typically deeply penetrating (also called
``horizontal'') or even Earth-skimming,
which can be used to distinguish them from cosmic and $\gamma-$ray
induced air showers. This is sketched in Fig.~\ref{fig:nu_auger}. The
tau-neutrino is especially suitable for Earth-skimming showers because
the tau-neutrino can produce a tau-lepton by a charged-current
interactions which travels with little energy loss before decaying
and inducing a giant air shower which can be detected. In this
detection mode, the decay length of the tau-lepton has to be
comparable to the dimensions of the detector, which limits sensitivity
to the energy interval between $\simeq10^{17}\,$eV and
$\simeq2\times10^{19}\,$eV in case of the Pierre Auger
Observatory~\cite{pierre_auger}. In a water or ice based detector such
as IceCube the neutrino charged current interaction can also be seen
which gives rise to ``double bang events''  at energies around
$10^{16}\,$eV.

From the non-observation of Earth-skimming
neutrinos the Pierre Auger experiment
has put an upper limit of $E_\nu^2j(E_\nu)\la30\,{\rm eV}\,{\rm cm}^{-2}\,{\rm
    sr}^{-1}\,{\rm s}^{-1}$ on the diffuse neutrino fluxes between
$\simeq10^{17}\,$eV and
$\simeq2\times10^{19}\,$eV~\cite{auger:2011pf,Abraham:2009uy}
that is comparable to the current IceCube
limit~\cite{Abbasi:2011ji} and also to the theoretical Waxman-Bahcall
and cascade limits, see Eq.~(\ref{eq:wb}) and~(\ref{eq:cascad}),
respectively, below. In the future, neutrino induced air
showers may also be observed from space which allows for a
considerable increase of effective target mass. The JEM-EUSO
experiment (``Extreme Universe Space Observatory onboard Japanese
Experiment Module'')~\cite{jem_euso} is considered to be flown on the
International Space Station (ISS). In the farther future one or
several freely flying satellites could be used for UHE cosmic ray and
neutrino detection.

We note in passing that the rate of horizontal neutrino induced air
showers is proportional to the product of the flux and the
neutrino-nucleon cross section because the atmosphere is transparent
even to the highest energy neutrinos. In contrast, the rate of Earth
skimming neutrinos tends to decrease with the neutrino-nucleon cross
section due to shadowing in the Earth crust. This may allow to break
the degeneracy of rates in fluxes and cross sections and thus would
allow to measure the UHE neutrino-nucleon cross section once a
sufficient number of both types of air showers are measured~\cite{Kusenko:2001gj}.

Finally, neutrinos can be detected from the radio emission of the
charged particle showers they produce in the atmosphere or at
surfaces of ice or rock: Since electrons created by ionization are
more mobile then the positive charges of the nuclei, there will be an
excess of electrons. In addition, the electrons interact differently
than nuclei and positrons. The resulting moving charge cloud will emit
coherent radio waves
either by interactions with the Earth's magnetic field, which is the
main emission mechanism in the atmosphere, or via Cherenkov emission,
known as the Askaryan effect, which dominates emission at ice and rock
surfaces. The latter effect has been used to set upper limits on the
UHE neutrino flux by several experiments such
as ANITA~\cite{anita} above $\simeq10^{17}\,$eV, a balloon flying
around the South Pole~\cite{Gorham:2010kv},
and by observing the moon's rim with radio telescopes such as the
Westerbork Synthesis Radio Telescope (the NuMoon
project)~\cite{Buitink:2010qn} above $\simeq10^{22}\,$eV. The use of
radio telescopes for observing radio pulses induced by neutrinos on
the moon's surface will become in particular interesting with the
arrival of Lofar~\cite{lofar} and SKA~\cite{ska} and the latter would
allow to lower the threshold to $\simeq10^{20}\,$eV. This technique can
also be used for UHE cosmic rays~\cite{terVeen:2010gb}. There are
further radio detection projects in Antartica, such as ARIANNA, ARA
and a ExaVolt Antenna (EVA)~\cite{Gorham:2011mt}, a next-generation
version of ANITA.

\section{High Energy Neutrino Fluxes}
\label{section7}
\begin{figure}[ht]
\includegraphics[height=0.9\textwidth,clip=true,angle=0]{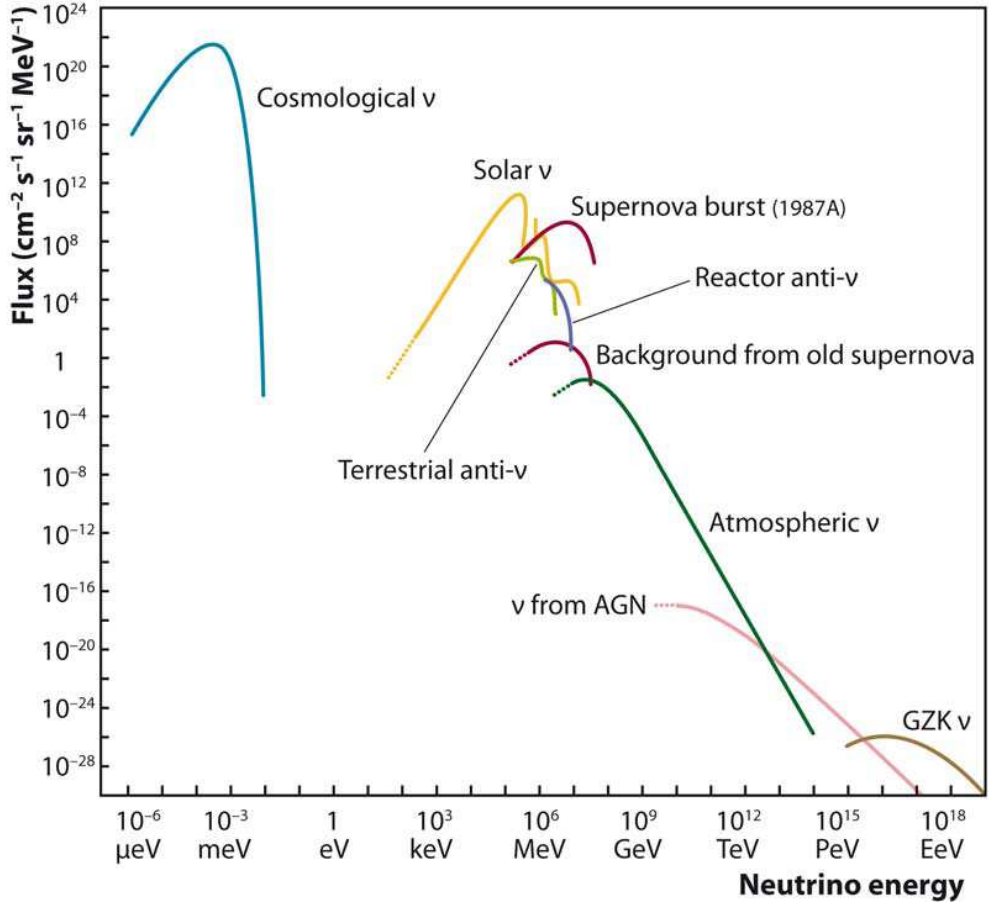}
\caption[...]{The diffuse ``grand unified'' neutrino spectrum. Solar
  neutrinos, a burst of neutrinos of a few seconds length
from SN1987A, reactor neutrinos, terrestrial neutrinos and atmospheric neutrinos have
already been detected. Another guaranteed although not yet detected flux is that of
neutrinos generated in collisions of ultra-energetic protons with the 3K cosmic
microwave background (CMB), the so-called GZK (Greisen-Zatsepin-Kuzmin)
neutrinos.  Whereas there is a good chance that GZK and AGN
neutrinos will be detected in the next decade, no practicable idea exists how to
detect 1.9 K cosmological neutrinos (the analogue to the 3K CMB). This
figure is taken from the ApPEC/ASPERA Astroparticle Physics Roadmap
Phase 1: ``Status and Perspectives of Astroparticle Physics in
Europe'' (Fig.4.1) at {\sl
  http://www.aspera-eu.org/images/stories/files/Roadmap.pdf}.}
\label{fig:nu-background}
\end{figure}

Fig.~\ref{fig:nu-background} shows the diffuse ``grand unified''
neutrino spectrum from the lowest energies corresponding to the cosmological relic
blackbody spectrum of temperature $\simeq1.9\,$K to the highest energy
neutrinos produced by interactions of UHECRs. In the following we will
restrict ourselves to the high energy range at a TeV and above.

The interactions of UHECR in general also give rise to secondary neutrino and
$\gamma-$ray fluxes. Therefore, the physics and astrophysics of UHECRs
is also inextricably linked with the emerging field of neutrino
astronomy~\cite{nu_review} and with the already well established field of $\gamma-$ray 
astronomy~\cite{gammarev}. In fact, neutrino and $\gamma-$ray
observations and limits often already severely constrain scenarios of UHECR origin.
In turn, this link plays an important role for theoretical
predictions of fluxes of extragalactic neutrinos with energies above
roughly a TeV whose detection is a major goal of several next-generation
neutrino telescopes including IceCube at the South pole and the
planned European KM3NeT: If such neutrinos are
produced as secondaries of UHECRs accelerated in astrophysical
sources and if these primary UHECRs are not absorbed within the sources,
but rather directly contribute to the observed UHECR flux, then
the energy fluence in the neutrino flux can not be higher
than the one in UHECRs. This corresponds to the so called Waxman-Bahcall
bound
\begin{equation}\label{eq:wb}
  E_\nu^2\frac{dN_\nu}{dE_\nu}\la(10-50)\,{\rm eV}\,{\rm
    sr}^{-1}\,{\rm s}^{-1}\,{\rm cm}^{-2}\,,
\end{equation}
where the range of values is due to uncertainties in the cosmological
evolution of the sources. The Waxman-Bahcall bound
applies to sources which are transparent to the high energy
cosmic rays they accelerate and which have acceleration spectra not much
harder than $E^{-2}$~\cite{wb-bound,mpr}. 
It does not hold if one of these assumptions
does not apply. This includes cases such as acceleration
sources with injection spectra considerably harder than $E^{-2}$,
so called ``hidden'' sources which are opaque to the primary UHECRs
that they accelerate, and top-down scenarios where UHECRs are produced
by very heavy relic particles decaying mostly into $\gamma-$rays and
neutrinos and only to $\sim10$\% into nucleons, rather than by
acceleration. In such cases where the Waxman-Bahcall bound does not
apply, the neutrino fluxes are still constrained by the observed diffuse $\gamma-$ray
flux in the 100 GeV range. This is because neutrinos are predominantly
produced through the decay of charged pions, due to approximate isospin
symmetry of known pion production channels roughly equal numbers of
positive, negative and neutral pions are produced. Therefore, a
comparable amount of energy goes into neutrinos and the
electromagnetic channel. But whereas neutrinos are only subject to redshift
during propagation to Earth, electrons, positrons and $\gamma-$rays
can initiate an electromagnetic cascade provided there is a sufficient
number of target photons on which the $\gamma-$rays can pair
produce. In an electromagnetic cascade the original $\gamma-$ray first produces
a pair on the low energy target photons where either the electron or
the positron takes most of the original energy and re-creates a
slightly lower energy $\gamma-$ray by inverse Compton scattering. This
newly created $\gamma-$ray starts a new cycle. The cycles continue to
shift the particle energies to lower values until the $\gamma-$ray energy falls below the
threshold for pair production on the low energy target photons. As a
consequence, the $\gamma-$ray flux piles up right below the pair
production threshold with a characteristic tail extending down to
lower energies. Such electromagnetic cascades can occur either within
individual UHECR sources on a target of infrared and optical
photons, or during propagation from the source to the observer, in
which case the main target will be the CMB and, to a lesser extent,
the large scale infrared background. In the latter case the Universe
thus acts like a calorimeter for electromagnetic energy injected above
the pair production threshold on the CMB,
$\sim10^{15}\,$eV. Electromagnetic energy injected above this threshold is
re-processed to a diffuse extragalactic $\gamma-$ray background at
hundreds of GeV. Recently, the higher sensitivity and better angular
resolution of the Fermi LAT experiment has allowed to resolve a larger
fraction of this background~\cite{Abdo:2010nz} originally measured by the EGRET
experiment~\cite{Strong:2003ex}. As a consequence, the best estimate
of the true diffuse extragalactic $\gamma-$ray background has
decreased. This has strengthened the so called cascade bound on the diffuse cosmic
neutrino flux~\cite{Berezinsky:1975zz} which now reads
\begin{equation}\label{eq:cascad}
  E_\nu^2\frac{dN_\nu}{dE_\nu}\la100\,{\rm eV}\,{\rm
    sr}^{-1}\,{\rm s}^{-1}\,{\rm cm}^{-2}
\end{equation}
and is thus already comparable to the original Waxman-Bahcall bound
Eq.~(\ref{eq:wb}). The cascade bound is also comparable to current experimental
sensitivities~\cite{Berezinsky:2010xa,Ahlers:2010fw}.

We now turn to one of the first scenarios for neutrino production
within discrete sources: Within the ``proton blazar'' model, in which AGNs are hadronic
accelerators, one can make a rough estimate of the diffuse neutrino
flux as contributed by the cosmological distribution of all proton
blazars as follows~\cite{Halzen:1997hw}:

\begin{itemize}

\item The size of the accelerator is $R\sim\Gamma T$, where the jet
  boost factor is $\Gamma\sim10$ and the duration of observed bursts
  is $T\sim1\,$day.

\item The magnetic field strength in the jet is estimated by
  $B^2\sim\rho_{\rm electron}\sim1\,{\rm erg}\,{\rm cm}^{-3}$ from
  equipartition with the plasma.

\item The ``Hillas condition'' Eq.~(\ref{eq:E_max}) on the maximal proton
  energy for a relativistic shock is $E_{\rm max}\sim eBR$
and from $p\gamma\to N\pi$ kinematics
$E_{\rm max,\nu}\sim0.1E_{\rm max}\sim10^{18}\,$eV.

\item The neutrino luminosity is related to the $\gamma-$ray luminosity by
$L_\nu\simeq\frac{3}{13}L_\gamma$ from $p\gamma\to N\pi$ kinematics.

\item Assume a differential proton spectrum $\Phi_p(E)=dN_p/dE\propto E^{-2-\beta}$
and a differential $\gamma-$ray spectrum $n_\gamma(\varepsilon)=dN_\gamma/d\varepsilon\propto
\varepsilon^{-2-\alpha}$.
If the jet is optically thin against $p\gamma\to N\pi$ then
$$
  \frac{dN_\nu}{dE_\nu}\propto \frac{dN_p}{dE}(10E_\nu)
  \int_{\varepsilon_{\rm N\pi}}^\infty d\varepsilon
  \frac{dN_\gamma}{d\varepsilon}\propto E_\nu^{-2-\beta}
  \left(\varepsilon_{\rm N\pi}\right)^{-1-\alpha}
  \propto E_\nu^{-1-\beta+\alpha}\,,
$$
since the pion production threshold
$\varepsilon_{\rm N\pi}\propto E^{-1}\propto E_\nu^{-1}$.

\item Combine this with the normalization,
$$
  \frac{dN_\nu}{dE_\nu}\simeq\frac{3}{13}\frac{L_\gamma}{E_{\rm max,\nu}}
  \frac{1-\beta+\alpha}{E_\nu}\,
  \left(\frac{E_\nu}{E_{\rm max,\nu}}\right)^{-\beta+\alpha}\,.
$$

\item Fold with the luminosity function of AGNs in GeV $\gamma-$rays.

\end{itemize}

The resulting diffuse energy fluence per energy decade
$E_\nu^2(dN_\nu/dE_\nu)$ peaks around $E_\nu\sim10^{17}\,$eV with a
maximum of $E_\nu^2(dN_\nu/dE_\nu)\simeq10^{-6}\,{\rm GeV}\,{\rm
  cm}^{-2}\,{\rm s}^{-1}\,{\rm sr}^{-1}$~\cite{Halzen:1997hw}
This optimistic model is already ruled out by the latest IceCube
limits which reads $E_\nu^2(dN_\nu/dE_\nu)\la10^{-7}\,{\rm GeV}\,{\rm
  cm}^{-2}\,{\rm s}^{-1}\,{\rm sr}^{-1}$ for $10^{15}\,{\rm eV}\la
E_\nu\la10^{18}\,$eV~\cite{Abbasi:2011ji}. Other blazar models have
also been ruled out already~\cite{Stecker:2005hn}.

Scenarios in which the sources of UHECR are primarily
GRBs~\cite{Waxman:1995vg,Dermer:2010km}
start to be strongly constrained by the non-observation so far of high
energy neutrinos which are predicted by these scenarios, both from
individual GRBs~\cite{Abbasi:2011qc} and from the accumulated diffuse flux from all
cosmological GRBs~\cite{Ahlers:2011jj}. Since the GRB environment is
sufficiently dense to confine the accelerated protons and nuclei, only
neutrons produced via $p\gamma\to n\pi^+$ can contribute to the UHECR
flux. This inevitably implies production of PeV scale neutrinos from
the decay of the pions produced in this reaction whose fluxes have
been computed in some
details~\cite{Waxman:1997ti,Anchordoqui:2007tn}. Neutrinos can be
emitted during the precursor phase, when the jet is still forming and
is optically thick for electromagnetic emission, during the
prompt phase when the GRB releases most of its $\gamma-$ray
emission, and during the afterglow phase.

The neutrons escaping from the GRB whose $\beta-$decay back
into protons which give rise to the UHECR flux $\Phi_p(E)$ per energy
from the GRB. Therefore, the neutrino per energy flux from the
interaction of the original accelerated proton, $p\gamma\to n\pi^+$,
can be directly related to $\Phi_p(E)$ via
\begin{equation}\label{eq:Phi_nu_GRB}
  \Phi_\nu(E_\nu)\sim\frac{1}{\eta_\nu}\Phi_p\left(\frac{E}{\eta_\nu}\right)\,,
\end{equation}
where $\eta_\nu\simeq0.1$ is the average neutrino energy in units of
the parent proton energy. This applies as long as the pions and muons
decay faster than they loose energy by interactions within the
GRB. Above a certain critical energy $E_{\pi,\mu}^c$ which is typically of the order
of $10^{17}\,$eV, this is not the case any more because of the
prolonged pion and muon lifetimes due to time dilation and the neutrino flux
above that energy is suppressed by a factor $\simeq
E_{\pi,\mu}^c/(4E_\nu)$ compared to Eq.~(\ref{eq:Phi_nu_GRB}) which
reflects the probability that the pion or muon decays before interacting.

The diffuse high energy neutrino flux is then obtained by folding
Eq.~(\ref{eq:Phi_nu_GRB}) with the cosmological GRB rate and
integrating over redshift. In general, the resulting
flux predictions are higher than current flux limits, as has been
shown for the neutrino flux associated with the prompt phase in
fireball scenarios in which associated UHECR fluxes are matched with
observations, for example, in Ref.~\cite{Ahlers:2011jj}. This does, however, not rule
out scenarios in which most of the UHECRs are produced by GRBs,
because the amount of energy transferred from protons accelerated in
the GRBs is uncertain by at least factors of a few.

We end with some remarks on the so called cosmogenic or GZK neutrino
flux which is produced by pion production of UHECR on the CMB during
propagation from the source to the observer. The flux of secondary cosmogenic
neutrinos~\cite{Anchordoqui:2007fi,Ahlers:2010fw,Kotera:2010yn}
and photons~\cite{Gelmini:2007jy,Hooper:2010ze} can in principle also
probe the UHECR mass composition: Secondary $\gamma-$rays and neutrinos are essentially
produced by pion production on the constituent nucleons of a
nucleus with a given atomic number $A$, as we have seen in
section~\ref{section4}. Therefore, if the maximal
acceleration energy $E_{\rm max}$ is not much larger than
$10^{21}\,$eV then for mass numbers $A$ approaching iron group nuclei,
the energy of the constituent nucleons will be below the GZK threshold
for pion production on the CMB. In this case, secondary $\gamma-$ray and neutrino
production can only occur by interactions with the infrared
background. Compared to pion production on the CMB, the rate of
pion production on the infrared background is
suppressed by the relative target photon number density
which is a factor of a few hundred. As a result, the cosmogenic
neutrino and photon fluxes depend strongly on injection spectrum,
maximal acceleration energy and mass composition. In general it
will not be easy to break the resulting degeneracies.


\acknowledgments
This work was
supported by the Deutsche Forschungsgemeinschaft through the
collaborative research centre SFB 676 Particles, Strings and the Early
Universe: The Structure of Matter and Space-Time and by the State of
Hamburg, through the Collaborative Research program Connecting
Particles with the Cosmos within the framework of the
Landesexzellenzinitiative (LEXI).



\bibliographystyle{aipproc}   


\IfFileExists{\jobname.bbl}{}
 {\typeout{}
  \typeout{******************************************}
  \typeout{** Please run "bibtex \jobname" to optain}
  \typeout{** the bibliography and then re-run LaTeX}
  \typeout{** twice to fix the references!}
  \typeout{******************************************}
  \typeout{}
 }

\end{document}